\documentclass[12pt, a4paper]{article}

\pdfoutput=1
\RequirePackage[numbers]{natbib}

\usepackage{amsthm,amsmath,natbib}

\usepackage{multirow}

\usepackage{listings}
\usepackage[latin1]{inputenc}
\usepackage[T1]{fontenc}
\usepackage[english]{babel}

\usepackage{subcaption}

\usepackage{articlestyle}
\usepackage{algorithm}
\usepackage{authblk}

\usepackage{fullpage}

\usepackage{ifpdf}
\usepackage{units}

\usepackage{mathtools}
\mathtoolsset{showonlyrefs}

\usepackage{xr}

\newcommand{\bs}[1]{\boldsymbol{{#1}}}

\DeclareMathOperator{\skp}{\gamma}
\DeclareMathOperator{\lkp}{\mu}
\DeclareMathOperator{\kp}{\tau}
\DeclareMathOperator{\prp}{\alpha}
\DeclareMathOperator{\spp}{\beta}

\DeclareMathOperator{\pNIG}{\proper{NIG}}
\DeclareMathOperator{\pIG}{\proper{IG}}
\DeclareMathOperator{\pGH}{\proper{GH}}
\DeclareMathOperator{\pGIG}{\proper{GIG}}

\title{Whole-brain substitute CT generation using Markov random field mixture models}
\author[1]{Anders Hildeman}
\author[1]{David Bolin}
\author[2]{Jonas Wallin}
\author[3]{Adam Johansson}
\author[4]{Tufve Nyholm}
\author[4]{Thomas Asklund}
\author[5]{Jun Yu}

\affil[1]{Department of mathematical sciences, Chalmers University of Technology and University of Gothenburg, Sweden}
\affil[2]{Department of Statistics, Lund University, Sweden}
\affil[3]{Department of Radiation Oncology, University of Michigan, USA}
\affil[4]{Department of Radiation Sciences, Umeå University, Sweden}
\affil[5]{Department of Mathematics and Mathematical Statistics, Umeå University, Sweden}

\begin{document}

\maketitle

\setcounter{tocdepth}{2}

\begin{abstract}
Computed tomography (CT) equivalent information is needed for attenuation correction in PET imaging and for dose planning in radiotherapy. Prior work has shown that Gaussian mixture models can be used to generate a substitute CT (s-CT) image from a specific set of MRI modalities. This work introduces a more flexible class of mixture models for s-CT generation, that incorporates spatial dependency in the data through a Markov random field prior on the latent field of class memberships associated with a mixture model. Furthermore, the mixture distributions are extended from Gaussian to normal inverse Gaussian (NIG), allowing heavier tails and skewness.  The amount of data needed to train a model for s-CT generation is of the order of $10^8$ voxels. The computational efficiency of the parameter estimation and prediction methods are hence paramount, especially when spatial dependency is included in the models. A stochastic Expectation Maximization (EM) gradient algorithm is proposed in order to tackle this challenge. The advantages of the spatial model and NIG distributions are evaluated with a cross-validation study based on data from 14 patients. 
The study show that the proposed model enhances the predictive quality of the s-CT images by reducing the mean absolute error with $17.9\%$. Also, the distribution of CT values conditioned on the MR images are better explained by the proposed model as evaluated using continuous ranked probability scores.
\end{abstract}

\section{Introduction}
\label{sec:introduction}
Ionizing radiation undergo attenuation as it passes through organic tissue. That
attenuation affects the dose deposition in radiotherapy and the image acquisition in positron emission tomography (PET). In both cases, the attenuation has to be estimated. 
The simulation of the dose distribution in radiotherapy makes it possible to optimize the treatment for the individual patient, maximizing the dose to the tumor while keeping the dose to healthy surrounding tissue within acceptable limits.
In PET, knowledge of the attenuation in the patient is a prerequisite for accurate quantification of the tracer uptake.

Computed Tomography (CT) X-ray imaging uses the attenuation of X-rays in order to construct a three dimensional image of the interior of the region of interest. Therefore, patients usually undergo a CT scan before radiotherapy treatment or in connection to the PET scan in order to acquire information about the attenuation. It has been shown that it is possible to derive similar attenuation information by the use of magnetic resonance imaging (MRI)  \citep{lit:adamCT, lit:sjolund}. Acquiring such a substitute CT (s-CT) image without exposing the patient to X-ray radiation has some advantages compared to performing a CT scan. Firstly, MRI does not expose the subject to ionizing radiation, which has an inherent risk of damaging tissue. Secondly, MRI information is often of interest for other reasons, for instance in order to increase the soft-tissue contrast and perform motion correction in PET images \citep{lit:BezrukovPETMRI}.

MR images do not map to CT images directly so in order to generate a s-CT image some prediction method is needed. There are two categories of methods, Atlas based and machine-learning based \citep{lit:BezrukovPETMRI}. The Atlas based methods finds a geometrical mapping between the MR images of the subject and those of MR images in a template library using image registration techniques. From the templates, CT images are then inversely mapped back to the subject and fused to give a s-CT image. The machine-learning based methods instead learn a mapping between the intensity values in the MR and CT images. This mapping is learned on training data where both MR and CT information is available. The mapping can then be applied for construction of s-CT images based solely on some corresponding MR images.

\citet{lit:adamCT} took the machine-learning based approach to this problem and utilized Gaussian mixture models (GMM) to map between MR and CT images. The parameters of the model were estimated using an Expectation-Maximization (EM) algorithm \citep{lit:EMAlgorithm}, and the s-CT images were constructed using the expected value of the CT field conditioned on the available MR images.  This method has the advantage that it is not dependent on an image registration step which could compromise prediction results. The model is also quite general and easy to estimate and has been shown to provide accurate results both for dose calculation in radiotherapy and attenuation correction in PET imaging \citep{lit:accuracypet, lit:accuracyradiotherapy}. 

The voxel values of the CT- and MR-images does not only depend on each other pointwise, there are also spatial dependencies that should be taken into account in a statistical model. \citet{lit:adamSpat} added a spatial component to the GMM approach by incorporating the spatial coordinates of the voxels as auxiliary dimensions of the data. Through this, each mixture class was given a spatial location. The spatial model showed improvements in the post-nasal cavities and inner ear where there is air and bone tissue in close proximity to each other.  However, a problem with giving mixture classes a spatial location is that areas separated in space but of the same tissue type needs to be modeled by different classes. This might yield problems with overfitting and unstable estimates due to the increasing demand for training data. Furthermore, the model does not make use of any spatial interaction between voxels and the mapping of the coordinates has the same drawbacks as Atlas based techniques, i.e. sensitivity to misregistration and abnormal anatomies. 

In this paper we take another approach to modeling the spatial dependency. A spatial Markov random field (MRF) model \citep[Chapter 4]{lit:Winkler} is applied as a prior distribution for the latent field of class memberships. This will bias voxel classification towards spatial clustering in a way that conveys local spatial structures without putting any global restrictions on the spatial location of the class distributions. Furthermore, the mixture model is extended by using a multivariate normal inverse Gaussian distribution (NIG) \citep{lit:Barndorff-NielsenNIG} for the class distributions. The NIG distribution adds flexibility since it allows for skewness and variable kurtosis which might reduce the number of classes needed to model heavy tailed or skewed data. 

A problem with these more flexible models is that they are computationally more demanding to estimate than the GMM.  Using models that incorporates spatial dependency for large datasets is not easy, and this is commonly referred to as the ``Big N'' problem in spatial statistics. In spatial statistics, datasets are usually considered to be big if they contain more than $10^4$ measurements. In this application, each image consists of more than 7 million voxels and during the learning phase five such images are needed per patient. Furthermore, a number of patients should be used in order to acquire reliable prediction parameters and all voxels need to be processed in each iteration of the learning algorithm. Thus, computational efficiency of the proposed methods is paramount. Here we introduce a novel approach utilizing the EM gradient algorithm \citep{lit:GradientEMAlgorithm} and Gibbs sampling to successfully and efficiently estimate parameters for mixture models including spatial dependency even in data rich environments such as this.

The remainder of this paper is divided in to four sections. Section \ref{sec:model} describes the MRF models and Section \ref{sec:parameterEstimation} introduces the method proposed in order to estimate those models. Section \ref{sec:results} describes a cross-validation study based on data from 14 patients, where the proposed models are compared with the original GMM. Results show that the new spatial model increases the predictive quality in comparison with the original model. Finally, Section \ref{sec:discussion} presents the conclusions and a discussion of future work. There is also an appendix giving some further details and derivations.

\section{Statistical modeling of CT-MR interdependence}\label{sec:model}
In order to find and make use of the dependency between a CT image and the corresponding MR images we will assume a parametric model. Since the data we consider consists of three-dimensional bit-mapped digital images one can consider each voxel (three-dimensional pixel) as a point on a three-dimensional equally spaced lattice. Let us enumerate these voxels $i \in \{1, ..., N\}$. We model the voxel values $\mathbf{X} = \{ \bs{X}_1, ..., \bs{X}_N \}$ as a random field on this discrete grid. Furthermore, since the data consists of one CT images and four MR images, each voxel value is five-dimensional, i.e. each $\mathbf{X}_i$ is five-dimensional.

\subsection{Mixture model}\label{sec:mixtures}
This paper extends the work of \citet{lit:adamCT} which used a GMM in order to model the interdependence between the MR images and the CT image. The probability density function of a general mixture model on $\mathbb{R}^d$ is $f( \mathbf{x}_i ) = \sum_{k=1}^K f_k (\mathbf{x}_i) \pi_k$, where $f_k$ is the density function of the distribution associated to class $k$, $\pi_k$ is the prior probability that $\mathbf{X}_i$ belongs to class $k \in \{1, ..., K\}$ where $K$ is the number of classes. A GMM is obtained if the $f_k$ are chosen as Gaussian, which is the most common choice in the literature.  

Let us denote the CT value of voxel $i$ as $X_i^{A}$ and a vector of the four MR values for voxel $i$ as $\bs{X}_i^{B}$. We will model the voxel values for all five images jointly as $\mathbf{X}_i = [X^A_{i}, \mathbf{X}^B_{i}]$ with a mixture model. Furthermore, we denote the set of the whole random field instead of just a single voxel by omitting the voxel index $i$, such as $\{\bs{X}_{i}\}_i = \mathbf{X} = [\bs{X}^A, \mathbf{X}^B]$.

Constructing a s-CT image from the MR images is equivalent to acquiring a prediction of the CT image from a realization of the random field $\mathbf{X}^B$. One predictor that we will use is the conditional expectation $\mathbb{E}[X_i^A|\mathbf{X}^B]$, which for a mixture model is
\begin{equation}
\mathbb{E}\left[ X_i^A | \mathbf{X}^B \right] =  \sum_{k=1}^K  \mathbb{E}[X_i^A|\mathbf{X}^B , Z_i = k] \mathbb{P}(Z_i = k | \mathbf{X}^B).
\label{eq:conditionalExpectationGeneralMixture}
\end{equation}
Here $Z_i \in \{1, ..., K\}$ is a latent variable that describes which mixture class voxel $i$ belongs to. 
As a measure of uncertainty of the prediction, we will use the conditional covariance,
\begin{align}
\mathbb{C}[X_i^A&|\mathbf{X}^B] =  \mathbb{E}\left[ X_i^A \left( X_i^A \right)^T \middle| \mathbf{X}^B \right] - \mathbb{E}\left[ X_i^A \middle| \mathbf{X}^B \right] \mathbb{E}\left[ X_i^A \middle| \mathbf{X}^B \right]^T \\
=& \sum_{k=1}^K   \mathbb{E} \left[X_i^A \left( X_i^A \right)^T \middle| \mathbf{X}^B, Z_i = k \right] \mathbb{P} \left(Z_i = k \middle|\bs{X}^B \right)  \\
&  - \sum_{k=1}^K \sum_{l=1}^K \left( \mathbb{E} \left[X_i^A \middle| \mathbf{X}^B, Z_i = k \right] \mathbb{E} \left[X_i^A \middle| \mathbf{X}^B, Z_i = l \right]^T \mathbb{P}\left(Z_i = k \middle|\bs{X}^B \right) \mathbb{P}\left(Z_i = l \middle|\bs{X}^B \right) \right) .
\label{eq:conditionalCovarianceGeneralMixture}
\end{align}

\subsection{Multivariate normal inverse Gaussian distribution}\label{sec:NIG}
In order to achieve a model that is more flexible than a GMM, the Gaussian class distributions can be exchanged to something more flexible. In this work a multivariate generalization of the NIG distribution is used. Other mixture distributions with similar advantages have been proposed before, for instance the skewed t- and skewed normal-distributions \citep[and the references within]{lit:sharonLee}.

The probability density function of the NIG distribution is 
\begin{align}
f(\boldsymbol{x}) =  \frac{ \sqrt{ \kp |Q|} }{ (2\pi)^{(d+1)/2}  } \exp\left(  \left( \boldsymbol{x} - \boldsymbol{\lkp} \right)^T Q  \boldsymbol{\skp} + \sqrt{ 2 \kp }  \right) 2 K_{\nu}(\sqrt{ab}) \left( \frac{b}{a} \right)^{\frac{\nu}{2}},
\label{eq:densityNIG}
\end{align}
where $K_{\nu}$ is a modified Bessel function of the second kind, $\nu = -\frac{d+1}{2}$, $a = \boldsymbol{\skp}^T Q \boldsymbol{\skp} + 2$, $b = (\boldsymbol{x}-\boldsymbol{\lkp})^TQ(\boldsymbol{x}-\boldsymbol{\lkp}) + \kp $. Here, $\bs{\skp}$, $\bs{\lkp}$, $Q$ and $\kp$ are parameters of the distribution where $\bs\lkp$ is a $d$-dimensional location parameter, $Q$ is a $d\times d$-dimensional positive definite symmetric matrix defining the interdependence between the dimensions, $\kp$ is a positive scalar that parametrize the kurtosis and $\bs\skp$ is $d$-dimensional skewness parameter. 

A useful representation of the NIG distribution is that $\bs{X} \sim  \pNIG(\boldsymbol\lkp, Q, \boldsymbol\skp, \kp)$ if
\begin{align}
&\mathbf{X} = \boldsymbol\lkp + \boldsymbol\skp V +  \sqrt{V}Q^{-\frac{1}{2}} \boldsymbol Z   \\
&V \sim \pIG ( \kp , \sqrt{\frac{\kp}{2}} ) \\
&\mathbf{Z} \sim \pN( \boldsymbol{\lkp} = \bs{0}, \bs{I}) 
\label{eq:NIGDefined}
\end{align}
Where $\pIG$ denotes the inverse Gaussian distribution, see \ref{app:GIG}, and $\pN$ denotes the multivariate Gaussian distribution. Note that $\mathbf{X}|V$ is multivariate Gaussian with $\mathbb{E}[\mathbf{X}|V] = \boldsymbol\lkp + \boldsymbol\skp V$ and precision matrix $\frac{1}{V}Q$. In \eqref{eq:NIGDefined} the IG distribution is parametrized by only one parameter, $\kp$ in order to avoid overparametrization.

\begin{figure}
\centering
\includegraphics[height = 6 cm, keepaspectratio]{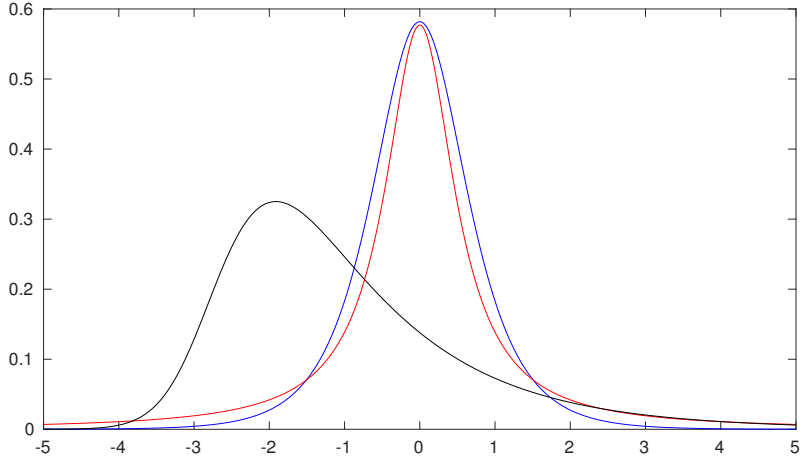}
\caption{Examples of probability density functions for three different set of values for the parameters of a univariate NIG distribution. }
\label{fig:NIG}
\end{figure}

In comparison to the Gaussian distribution, NIG is more flexible since it allows for arbitrary skewness and kurtosis. Also, the Gaussian distribution can be characterized as a special limiting case of the NIG distribution. 
Figure \ref{fig:NIG} shows three examples of density functions for a univariate NIG distribution.

In order to perform probabilistic prediction of CT images the marginal and conditional distributions need to be used. The following two propositions are therefore useful. 
\begin{prop}\label{prop:marginalNIG}
Suppose that $\bs{X}\sim\pNIG(\bs{\lkp}, Q, \bs{\skp}, \kp)$ and let $\bs{X} = \left[ \bs{X}^A, \bs{X}^B \right]^T$, then $\mathbf{X}^B \sim  \pNIG( \boldsymbol\lkp^B, (\Sigma^{BB})^{-1}, \boldsymbol{\skp}^B, \kp)$. Here $\boldsymbol\lkp^B$,   $\Sigma^{BB}$, and $\boldsymbol{\skp}^B$ are the parts of $\boldsymbol\lkp$,  $\Sigma$, and $\boldsymbol{\skp}$ respectively that correspond to $\bs{X}^B$.
\end{prop}

\begin{prop}\label{prop:conditionalNIG}
Suppose $\bs{X} \sim \pNIG(\bs{\lkp}, Q, \bs{\skp}, \kp)$, and let $\bs{X} = [\bs{X}^A, \bs{X}^B]^T$, then $\mathbf{X}^A | \mathbf{X}^B$ has density
\begin{equation}
f(\mathbf{x}^A | \mathbf{x}^B) = 
\frac{ K_{\nu} \left( \sqrt{ ab } \right) }{ K_{\hat{\nu}} \left( \sqrt{ \hat{a}\hat{b} } \right) } \left( \frac{\hat{a}}{\hat{b}} \right)^{\hat{\nu}/2} \left( \frac{b}{a} \right)^{\nu/2} e^{ \left(  \boldsymbol{\skp}^T Q (\mathbf{x}-\boldsymbol{\lkp})  -\boldsymbol{\skp}^B \left(\Sigma^{BB}\right)^{-1} (\mathbf{x}^B - \boldsymbol{\lkp}^B) \right)},
\label{eq:conditionalDensityNIG}
\end{equation}
where $\hat{\nu} = -\frac{|B|+1}{2}$, $|B|$ is the dimensionality of $\bs{X}^B$, $\hat{a} = \left(\bs{\skp}^B\right)^T \left(\Sigma^{BB}\right)^{-1} \boldsymbol{\skp}^B + 2$ and $\hat{b} = (\mathbf{x}^B - \boldsymbol{\lkp}^B)^T \left(\Sigma^{BB}\right)^{-1} (\mathbf{x}^B - \boldsymbol{\lkp}^B) + \tau$. Further, the conditional mean and covariance are
\begin{align}
\mathbb{E}[\mathbf{X}^A | \mathbf{X}^B] &= \boldsymbol{\lkp}^A - \left(Q^{AA}\right)^{-1}Q^{AB} \left( \boldsymbol{x}^{B} - \boldsymbol{\lkp}^{B} \right)  \\
& +\left( \boldsymbol{\skp}^A + \left(Q^{AA}\right)^{-1}Q^{AB}  \boldsymbol{\skp}^{B} \right) \sqrt{\frac{\hat{b}}{\hat{a}}} \frac{K_{\hat{\nu}+1}(\sqrt{\hat{a}\hat{b}})}{K_{\hat{\nu}}(\sqrt{\hat{a}\hat{b}})}, \\
\mathbb{C}( \mathbf{X}^A | \mathbf{X}^B ) &=  Q_{A,A}^{-1} \sqrt{\frac{\hat{b}}{\hat{a}}} \frac{ K_{\hat{\nu}+1}(\sqrt{\hat{a}\hat{b}}) }{K_{\hat{\nu}}(\sqrt{\hat{a}\hat{b}})} \\ 
&+ \hat{\boldsymbol{\skp}}\hat{\boldsymbol{\skp}}^T \frac{\hat{b}}{\hat{a}}  \left[ \left( 2 \frac{\hat{\nu}+1}{\sqrt{\hat{a}\hat{b}}} -   \frac{ K_{\hat{\nu}+1}(\sqrt{\hat{a}\hat{b}}) }{K_{\hat{\nu}}(\sqrt{\hat{a}\hat{b}})}   \right)  \frac{ K_{\hat{\nu}+1}(\sqrt{\hat{a}\hat{b}}) }{K_{\hat{\nu}}(\sqrt{\hat{a}\hat{b}})}  
 + 1 \right].
\end{align} 
\end{prop}
For a proof of these propositions, see \ref{app:propNIG}. 
Note that $\mathbf{X}^A | \mathbf{X}^B$ is not NIG distributed but has as a generalized hyperbolic distribution \citep{lit:Hammerstein}. The generalized hyperbolic distribution generalizes NIG by letting the mixing variable, $V$, be distributed as a generalized inverse Gaussian distribution (GIG), see \ref{app:GIG} for a definition.

\subsection{Spatial dependency} \label{sec:spatialDependence}
In a structured image, such as an MR- or CT- image, the voxel values will not be independent. While still utilizing a mixture model, we infer spatial dependency by applying a spatially dependent prior on the class membership field, $\mathbf{Z} = \{Z_i\}_i$, where $Z_i$ indicates the class membership of voxel $i$. However, to simplify estimation, we still assume conditional independence between voxel values conditioned on $\bs{Z}$, i.e. $\bs{X}_i \perp \bs{X}_j | Z_i , Z_j$ \text{for} $i \neq j$.

The dependency structure in $\bs{Z}$ is modeled with an MRF on the three-dimensional lattice defined through the conditional probability in \eqref{eq:potts},
\begin{equation}
\mathbb{P}(Z_i = k | \mathbf{Z}_{-i} = \bs{z}_{-i}) = \frac{1}{W_i(\mathbf{Z}_{-i}, \bs{\prp}, \spp)} \exp\left(-\prp_k - \sum_{j \in \mathcal{N}_i} \spp_{k z_j} \right),
\label{eq:potts}
\end{equation}
where $\mathcal{N}_i$ is the set of all neighbors to the $i$:th voxel, $\prp_k$ is the unconditional probability potential of class $k$ and $\spp_{k l}$ is the conditional probability potential of class $k$ attributed to neighbors of class $l$. Further, $-i$ is used to denote the set of indices to all voxels except $i$, i.e. $-i = \{1, ..., N\} \setminus \{i\}$ and $\bs{Z}_{-i} = \{Z_j\}_{j\in -i}$. Finally $W_i(\mathbf{Z}_{-i}, \bs{\prp}, \spp) = \sum_{k} \exp\left(-\prp_k - \sum_{j \in \mathcal{N}_i} \spp_{k z_j} \right)$ is a normalizing constant.

The probability density function of $\bs{X}_i$ conditioned on the class identities of all other voxels is $f( x_i | \mathbf{z}_{-i}) = \sum_{k = 1}^K f_k( x_i ) \mathbb{P}( Z_i = k | \bs{Z}_{-i})$. Hence, the $\spp_{kl}$ parameters describe how classes attract (negative values) or repel (positive values) each other in the topological lattice space. 

\begin{figure}
\centering
\begin{subfigure}{0.45\textwidth}
\centering
\includegraphics[width = 4cm, keepaspectratio]{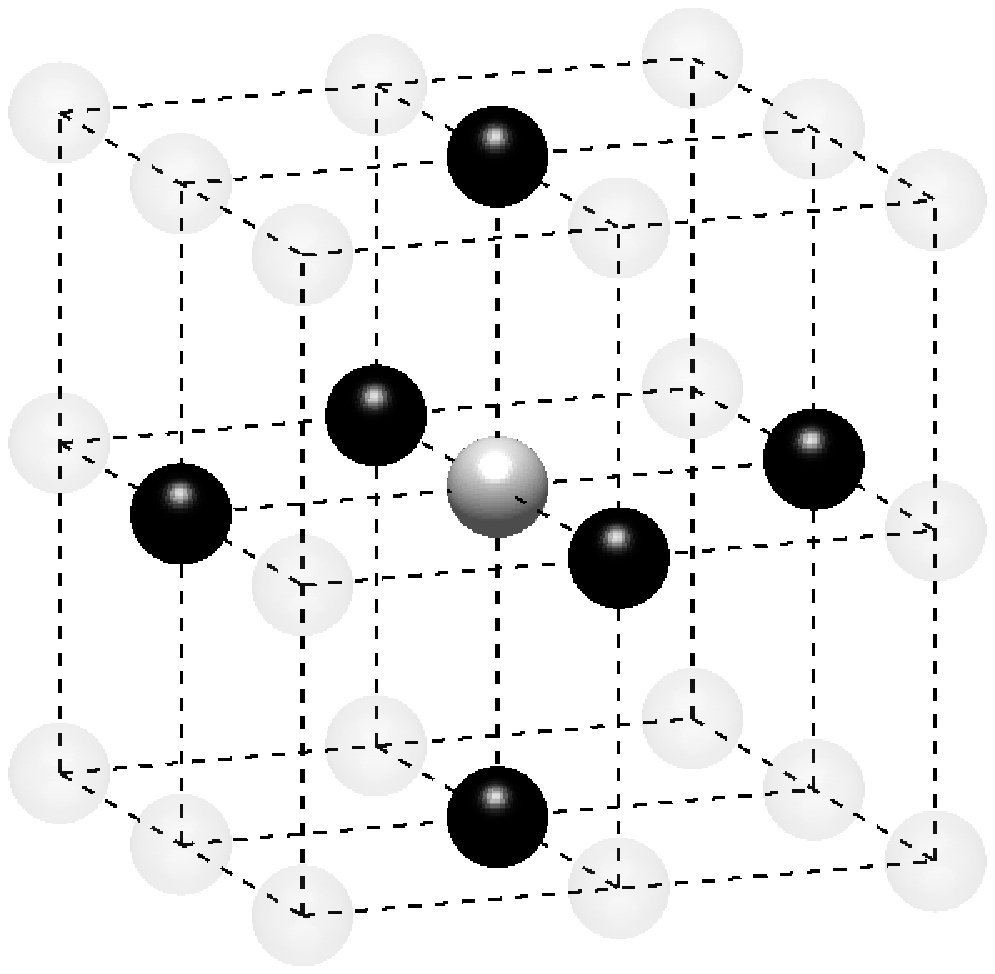}
\caption{}
\label{subfig:neigh}
\end{subfigure}
\begin{subfigure}{0.45\textwidth}
\centering
\includegraphics[width = 4cm, keepaspectratio]{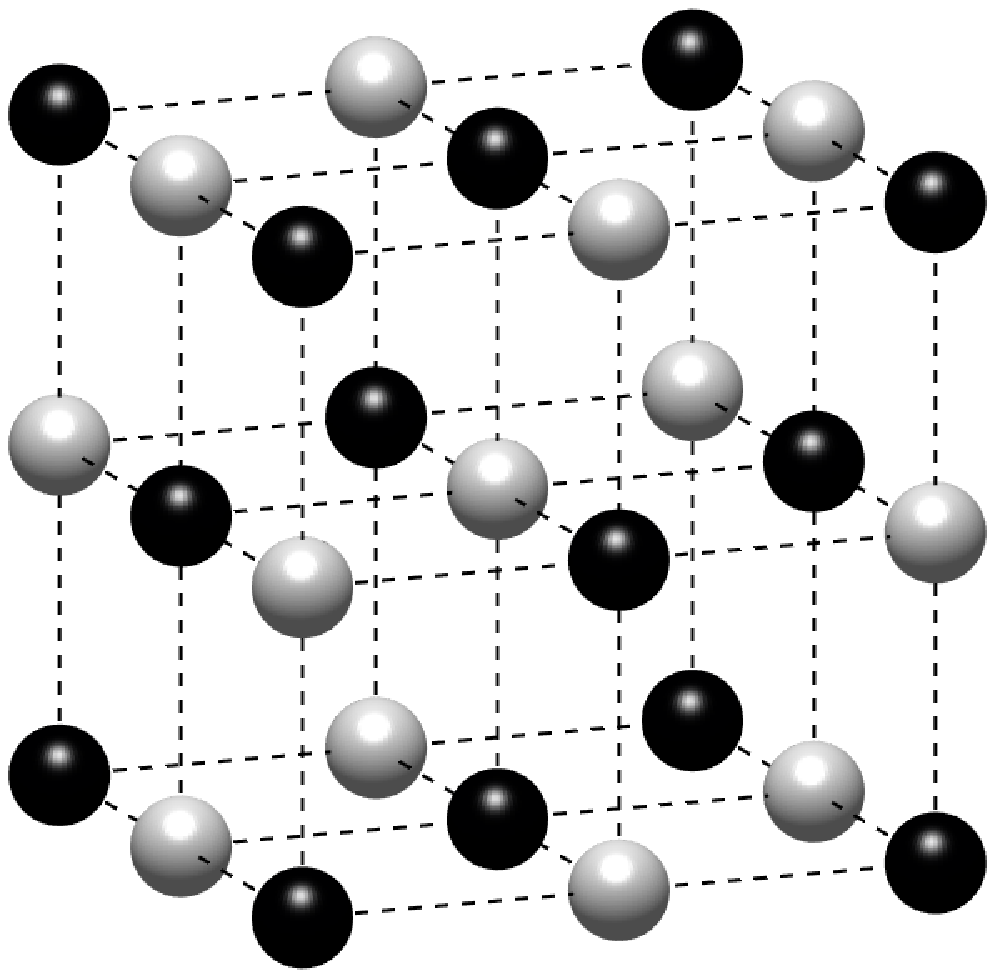}
\caption{}
\label{subfig:checkers}
\end{subfigure}
\caption{First-order neighborhood structure in three dimensions for an equidistant lattice. For any white ball the nearest neighbors will be black and vice verse}
\label{fig:neigh}
\end{figure}

Since the unconditional probability potentials, $\prp_k$, overparametrizes the conditional probability model by one degree of freedom, we let $\prp_1 = 0$ to make the model identifiable.
Furthermore, we choose a first-order neighborhood structure (the six nearest neighbors in three dimensions), see Figure \ref{fig:neigh}, and for simplicity we restrict the $\spp_{kl}$ values to 
\begin{equation}
\spp_{kl} = \begin{cases}
\begin{aligned}
0 &, k \neq  l \\
\spp &, k = l
\end{aligned}
\end{cases} .
\end{equation}
This means that there will only exist a conditional probability potential between neighboring voxels if they are of the same class and in this case the potential will be the same regardless of the class. This corresponds to the standard Potts model of \citet{lit:Wu}. Even though we in this paper restrict $\spp_{kl}$ to this simplified model, more general models with arbitrary $\spp_{kl}$ values can be estimated using the same theory and methods.

From here on, we will refer to a model using an MRF prior for the class memberships as a spatial model. 

\section{Parameter estimation and prediction}\label{sec:parameterEstimation}
In order to use the models in Section \ref{sec:model} for s-CT generation the model parameters need to be estimated from data. We choose to do this using a maximum likelihood approach. The likelihood function is
\begin{align}
L(\Theta ; \bs{x}) &=  \sum_{\bs{z} \in \Omega(\bs{Z})} f(\bs{x}| \bs{z}; \Theta) \mathbb{P}(\bs{Z} = \bs{z} ; \Theta)  = \sum_{\bs{z} \in \Omega(\bs{Z})} \left( \prod_i f(\bs{x}_i| z_i ; \Theta) \right) \mathbb{P}(\bs{Z} = \bs{z} ; \Theta),
\label{eq:Likelihood}
\end{align}
where $\Theta$ is a set of parameter values for the model, $\bs{x}$ is a realization of the voxel field $\bs{X}$ described in Section \ref{sec:mixtures} and $\Omega(\bs{Z})$ is the finite sample space of $\bs{Z}$, i.e. the set of all $K^N$ possible combinations of class identities for the voxels.

Due to the Hammersley-Clifford theorem we know that our MRF is a neighbor Gibbs field for the first order neighborhood structure \citep[Chapter 4]{lit:Winkler}. Hence, there exists a closed form expression, 
\begin{align} \label{eq:Gibbs}
\mathbb{P}(\bs{Z} = \bs{z}) = \frac{1}{W(\bs{\prp}, \spp)}\exp\left( - \sum_{i = 1}^N \left( \alpha_{z_i}  + \frac{1}{2}\sum_{j \in \mathcal{N}_i} \mathbb{I}_{z_i = z_j}\beta \right)  \right).
\end{align}
For the spatial models, the partition function, $W(\bs{\prp}, \spp)$, is unfortunately not feasible to compute since it requires summation over all possible states of $\bs{Z}$. Instead we replace $L$ with the pseudolikelihood  
\begin{align}
\tilde{L}(\Theta ; \mathbf{x}) &=  \sum_{\mathbf{z} \in \Omega(\bs{Z})} \prod_{i = 1}^N f(\bs{x}_i| z_i ; \Theta) \mathbb{P}(Z_i = z_i | \bs{Z}_{-i}=\bs{z}_{-i}; \Theta),    
\label{eq:pseudoLikelihood}
\end{align}
i.e.  the joint probability of $\bs{Z}$ is approximated as the product of all conditional probabilities. 

For a non-spatial model the pseudolikelihood is not an approximation since the class membership of the voxels are independent of each other. For a spatial model there is however a discrepancy  and approximating the joint distribution of $\bs{Z}$ as in \eqref{eq:pseudoLikelihood} can be motivated using the reasoning of \citep[Section 6.1]{lit:BesagInteractionLattice}.

\subsection{EM gradient algorithm}
\label{sec:EMGradient}
Commonly, maximum likelihood estimates of mixture models are acquired using the EM algorithm \citep{lit:EMAlgorithm}. This corresponds to iteratively finding $\Theta^{(j+1)} = \arg\max Q(\Theta | \Theta^{(j)})$ where $\Theta^{(j)}$ is a vector of the estimated parameter values at the $j$:th iteration and
\begin{align}
Q \left(\Theta | \Theta^{(j)} \right) &= \mathbb{E}_{\bs{Z}} \left[ \log L(\Theta | \bs{Z} , \bs{X} = \bs{x}) \middle| \bs{X} = \bs{x} ; \Theta^{(j)} \right] .
\label{eq:EMQFunc}
\end{align}

Performing the E-step corresponds to computing the posterior probabilities for each voxels class membership, i.e. $\{\mathbb{P}(Z_i = k | \mathbf{X} = \bs{x}; \Theta^{(j)}) \}_{i=1,k=1}^{N,K}$. For the non-spatial models, conditional class probabilities are simply 
$\mathbb{P}(Z_i = k | \bs{X} = \bs{x}, \Theta) = f_k(\bs{x})\pi_k/(\sum_l f_l(\bs{x})\pi_j)$.
For the spatial models, only the conditional probabilities $\mathbb{P}(Z_i = k | \mathbf{z}_{-i})$ are known explicitly. However, the probability $\mathbb{P}(Z_i = k | \mathbf{x})$ can be estimated through Monte Carlo simulation since 
\begin{equation}\label{eq:posteriorLatentField}
\mathbb{P}(Z_i = k| \mathbf{X} ; \Theta^{(j)}) =  \mathbb{E}_{\bs{Z}} \left[ \mathbb{I}_{Z_i = k} \middle| \bs{X}  ; \Theta^{(j)} \right]. 
\end{equation}
Here, Gibbs sampling can be used to estimate the expectation, see \ref{app:MCMC} for details. 

Performing the M-step is straightforward for a GMM, but in general it is often difficult to derive explicit expressions for the updates. In particular, it is not computationally feasible to estimate the spatial models using a standard EM algorithm.  Using an EM algorithm with an iterated conditional modes (ICM) or some Markov Chain Monte Carlo (MCMC) based estimator for the spatial parameters would be a possibility and such methods have been used in similar applications \citep{lit:MRIMRFBrainSeg,lit:Zhang}. However, the computational burden increases significantly if iterative methods are used in each M-step and for the purpose of whole-brain s-CT generation we need a more computationally efficient estimation method.

If it is possible to evaluate the gradient of the log likelihood with regards to the parameters, $\nabla \log \tilde{L}(\theta; \bs{x})$, an alternative to the EM algorithm would be to use gradient-based optimization in order to maximize the likelihood. 
The gradient can be expressed as
\begin{align}
\nabla \log &L( \Theta ; \bs{x}) = \nabla \log f( \bs{x} ; \Theta ) = \frac{\nabla f( \bs{x} ; \Theta )}{f( \bs{x} ; \Theta )} = \frac{1}{f( \bs{x} ; \Theta )} \sum_{\mathbf{z}} \nabla f( \bs{x}, \mathbf{z} ; \Theta ) \\
&=  \sum_{\mathbf{z}} \frac{ f( \bs{x}, \mathbf{z} ; \Theta ) }{f( \bs{x} ; \Theta )} \nabla \log f( \bs{x}, \mathbf{z} ; \Theta ) = \sum_{\mathbf{z}} f( \mathbf{z} | \bs{x} ; \Theta ) \nabla \log f( \bs{x}, \mathbf{z} ; \Theta ) \\
&= \mathbb{E}_{\mathbf{Z}} \left[ \nabla \log f( \bs{x}, \bs{Z} ; \Theta ) | \bs{X} = \bs{x} ; \Theta \right] \\ 
&= \mathbb{E}_{\mathbf{Z}} \left[ \left( \nabla \log f( \bs{x} | \mathbf{Z} ; \Theta ) +  \nabla \log \mathbb{P}( \bs{Z} = \mathbf{z} ; \Theta ) \right) | \bs{X} = \bs{x} ; \Theta \right] .
\end{align}
Analogously, for the pseudolikelihood this yields
\begin{align}\label{eq:gradientLogPseudLik}
\nabla \log &\tilde{L}( \Theta ; \bs{x}) =  \sum_{i=1}^N\tilde{\mathbb{E}}_{\mathbf{Z}} \left[   \nabla \log f( \bs{x}_i | Z_i ; \Theta ) \middle| \bs{X} = \bs{x} ; \Theta \right]  \\
&+  \sum_{i=1}^N \tilde{\mathbb{E}}_{\mathbf{Z}} \left[\nabla \log \mathbb{P}( Z_i = z_i | \bs{Z}_ {-i} = \mathbf{z}_{-i} ; \Theta )  \middle| \bs{X} = \bs{x} ; \Theta \right] .
\end{align}
Where $\tilde{\mathbb{E}}$ denotes the expectation taken according to the probability distribution of $\bs{Z} | \bs{X}$ induced by the pseudolikelihood.
The expressions inside the expectation are all available in explicit form and it is possible to estimate the expectation using Gibbs sampling in the same way as was done for $\mathbb{P}(Z_i = k | \mathbf{x})$, see \ref{app:MCMC} for details.

Since we can obtain an approximation of the gradient of the log likelihood, we can apply a gradient ascent algorithm to estimate the maximum likelihood parameters of the model, and iteratively update $\Theta$ as follows until convergence,
\begin{align}
\Theta^{(j+1)} = \Theta^{(j)} + \delta^{(j)} \nabla \log \tilde{L}( \Theta^{(j)} ; \mathbf{x}) .
\end{align}
In this case, if $\delta^{(j)}$ is a sequence of step lengths with sufficiently small but positive values, $\{\Theta^{(j)}\}$ will converge to a stationary point of $\tilde{L}(\Theta ; \bs{x})$ if one exists and if $\tilde{L}$ is first order continuous and bounded. 

Note that it is possible to evaluate the gradient if one can evaluate the conditional class probabilities, $\mathbb{P}(Z_i = k | \bs{X} = \bs{x}; \Theta)$. Finding class probabilities is equivalent to finding the expected value of the latent variable $\hat{z}_{ik} = \mathbb{I}_{Z_i = k}$ and is hence equivalent to an E-step in the regular EM algorithm. Because of this, using these gradient-based methods corresponds to, in each iteration, performing an E-step followed by taking a step in parameter space to a new set of parameter values. Thus, the method can be viewed as an EM algorithm where the M-step is approximated by one step of a gradient-based optimization method.

It is generally hard to chose values of $\delta^{(j)}$ that lead to fast and reliable convergence. Moreover, choosing the parameter path of steepest gradient is often suboptimal. One option is to replace $\delta^{(j)}$ with some scaling matrix $S^{(j)}$ that leads to a more optimal parameter path. If $\tilde{L}$ is two times differentiable, a particular choice of $S^{(j)}$ is minus the inverse Hessian of the log likelihood. This choice of the $S^{(j)}$ matrix corresponds to Newtons method for finding zeros of $\nabla \tilde{L}$, and an update then looks like
\begin{align}
\Theta^{(j+1)} = \Theta^{(j)} - H^{-1}(\log \tilde{L}(\Theta^{(j)}; \bs{x})) \nabla \log \tilde{L}( \Theta^{(j)} ; \mathbf{x}) .
\label{eq:newtonGradient}
\end{align}
Newtons method has superlinear convergence rate in a concave neighborhood to a stationary point \citep{lit:Bertsekas}. This in comparison to the linear convergence rate of general choices of positive definite $S^{(j)}$ matrices (with small enough eigenvalues) suggests that Newtons method should be used when applicable.

This particular algorithm where the approximate M-step is performed by one iteration of Newtons method is known as the EM gradient algorithm \citep{lit:GradientEMAlgorithm}.  
This is the main outline of our estimation method. However, some modifications are needed to ensure convergence for our problem. These modifications are presented briefly in the subsections below, and the resulting estimation method is outlined in Algorithm \ref{alg:estimation}.

\begin{algorithm}[t]
\begin{algorithmic}[1]
\Procedure{estimateParams}{$\bs{x}$,$\bs{z}$,$\Theta$}

\While{ step sizes large enough } 
\For{$k$ in $K$ } 
\State $\bs{p}_k = \alpha_k + \log f(\bs{x} ; \Theta_k)$	
\EndFor

\State $\{\bs{p}, \bs{z}, \bs{da}, \bs{db}\} =$ GibbsSample$( \bs{x}, \bs{z}, \bs{p} )$             \Comment{See \ref{app:MCMC}}

\State $ \bs{d}\Theta = $ ComputeGrad$( \bs{x}, \bs{p}, \Theta )$ \Comment{See section \ref{sec:EMGradient}}

\State $\bs{s} =$ EMGrad$( \Theta, \bs{d}\Theta, \bs{da}, \bs{db} ) $

\State $\bs{s} =$ LineSearch$( \Theta, \bs{s} )$ \Comment{See \ref{app:condLineSearch}}

\State $\Theta = \Theta + \bs{s}$

\EndWhile

\EndProcedure
\end{algorithmic}
\caption{Parameter estimation procedure.}
\label{alg:estimation}
\end{algorithm}

\subsubsection{Line search}\label{sec:condLineSearch}
Given that the Hessian matrix is negative definite for all iterations, choosing $S$ as the negative inverse Hessian matrix scaled by some small enough step length will lead to convergence of the parameter estimation. Instead of choosing a fixed scaling of the Hessian, one could do a line search in the direction given by Newtons method to find the best scaling in each iteration. This line search procedure corresponds to performing an improved approximate M-step in the EM-gradient method and is recommended in \citet{lit:GradientEMAlgorithm}. In fact, if the Hessian is negative definite and if a line search is performed, the EM gradient algorithm has a convergence rate of the same order as the regular EM algorithm.  

Since we do not have a closed-form expression for the likelihood, performing line search is not straightforward for the spatial models. Instead one can perform a line search conditioned on the E-step as explained in \ref{app:condLineSearch}.

\subsubsection{Sampling and conditioning of $S$}
A problem with using a scaled negative inverse Hessian as $S$ is that it is not guaranteed that the Hessian will be negative definite. When this assumption fails, Newtons method does not necessarily converge to a point in the parameter space with a higher likelihood than the initial value. However, by conditioning $S$ to be positive definite, and performing a line search, it is easy to see that $Q(\Theta^{(j+1)} | \Theta^{(j)}) \ge Q(\Theta^{(j)} | \Theta^{(j)})$, and the inequality will be strict as long as $\Theta^{(j)}$ is not a stationary point. To achieve a positive definite $S$ we first check if the computed Hessian is negative definite. If not, $S$ is chosen as the negative diagonal of the Hessian. If $S$ is still not positive definite, the diagonal entries are translated until all of them are positive. 

A method that iteratively performs an E-step followed by an approximate M-step that guarantees $Q(\Theta^{(j+1)} | \Theta^{(j)}) > Q(\Theta^{(j)} | \Theta^{(j)})$ is known as a Generalized EM algorithms (GEM) \citep{lit:EMAlgorithm}. Thus, by conditioning $S$ to be positive definite and performing a line search conditioned on the E-step, the proposed estimation algorithm belongs to the class of GEMs. Hence, it will always increase the likelihood and when the assumptions for Newtons method are fulfilled it will also  converge with super linear rate. 

The E-step for the spatial models is only approximate due to the MC sampling. So far we have not assessed if the Monte Carlo errors affects the convergence results. However, in analogy to the MCEM algorithm of \citet{lit:Wei}, the MCMC approximation in our E-step give us, not a GEM method, but a Monte Carlo GEM method. Convergence of such an algorithm follow analogously from the convergence of the MCEM algorithm \citep{lit:Chan}.

\subsubsection{Approximate Hessian}
\label{sec:approximateHessian}
The final modification that is needed to make the estimation method work is to approximate the Hessian. The reason for this is that it is not possible to estimate the true Hessian using Monte Carlo simulation in the same manner as was possible with the gradient. The Hessian can be written as 
\begin{align*}
H(\log &\tilde{L}( \Theta; \mathbf{x})) =  \sum_{i = 1}^N \tilde{\mathbb{E}}_{\mathbf{Z}} \left[  H \left( \log f( \mathbf{x}_{i} | Z_i ; \Theta )\right)  + H \left( \log f( Z_{i} | \mathbf{Z}_{-i} ; \Theta )\right) | \bs{X} = \mathbf{x} ; \Theta \right] \\
 &+ \sum_{i = 1}^N \tilde{\mathbb{E}}_{\mathbf{Z}} \left[   \nabla \log f( \mathbf{x}_{i} | Z_{i} ; \Theta )\nabla \log f(\mathbf{Z} | \bs{X} = \mathbf{x} ; \Theta)^T | \bs{X} = \mathbf{x} ; \Theta \right] \\
 &+ \sum_{i = 1}^N \tilde{\mathbb{E}}_{\mathbf{Z}} \left[ \nabla \log f( Z_{i} | \mathbf{Z}_{-i} ; \Theta )  \nabla \log f(\mathbf{Z} | \bs{X} = \mathbf{x} ; \Theta)^T | \bs{X} = \mathbf{x} ; \Theta \right]. 
\end{align*}
Here, the last two terms include $\nabla \log f(\mathbf{Z} | \bs{X} = \mathbf{x} ; \Theta)^T$ which we do not have a closed form expression for. Instead, we will approximate the Hessian using only the first term, in hope that this term dominates the Hessian. This approximation highlight yet another reason why it is necessary for us to perform a proper line search and conditioning of $S$ to assure convergence.

\subsection{CT prediction}
Given parameter estimates based on training data, the conditional expectation from Equation \eqref{eq:conditionalExpectationGeneralMixture} can be used to generate the s-CT images for a new patient where only MR images are recorded. Thus, the CT value for each voxel is predicted using the formula
\begin{align}
\mathbb{E} [X_{i}^A | \bs{X}^{B} = \bs{x}^{B}] = \sum_{k=1}^K \mathbb{E}_k[X_{i}^A | \bs{X}_{i}^B = \bs{x}_{i}^B] \mathbb{P}( Z_i = k | \bs{X}^{B} = \bs{x}^{B} ),
\end{align}
where $X_{i}^A$ is the CT value of the $i$:th voxel and $\bs{X}_{i}^B$ are the MR values for the same. Here, $\mathbb{E}_k[X_{i}^A | \bs{X}_{i}^B=\bs{x}_{i}^B]$ has an analytical closed form expression and $\mathbb{P}( Z_i = k | \bs{X}^{B}=\bs{x}^{B} )$ can be approximated using MCMC simulation analogously to how $\mathbb{P}(Z_i = k| \bs{X}=\bs{x} )$ was approximated, see \ref{app:MCMC}.

Using the conditional mean as the predictor of the s-CT image corresponds to minimizing the root mean square error (RMSE) of the prediction, based on the model assumption being correct. If one instead would be interested in minimizing the mean absolute error (MAE) the conditional median would be a more appropriate predictor \citep{lit:gneiting}. There is no analytical expression for the conditional median but given $\mathbb{P}( Z_i = k | \bs{X}^{B}=\bs{x}^{B} )$ it can easily be approximated by Monte-Carlo simulation since the conditional distribution is known. 

\subsection{Computational cost}
In \ref{app:MCMC} it is shown that the computational complexity of the MCMC sampling used to approximate the expectations in the EM gradient method is of order $\mathcal{O}(J N K )$, where $J$ is the number of Monte Carlo iterations, $K$ is the number of classes and $N$ is the number of voxels in the image. Here, $J$ can be chosen low since MCMC chains of consecutive EM-gradient iterations can feed of the former to reduce burnin. For the results presented later, $J = 10$ was found to be sufficient. 
Besides the MCMC sampling, each iteration of the algorithm requires summing up computations voxelwise as well as classwise. Hence, the computational complexity is $\mathcal{O}(n J N K)$, where $n$ is the number of EM gradient iterations.

Analogously, once the model parameters are available, the CT prediction has a computational complexity of $\mathcal{O}(J N K)$ due to the need to generate a MCMC chain for approximating $\mathbb{P}( Z_i = k | \bs{X}^{B}=\bs{x}^{B} )$. For the prediction step, $J$ needs to be larger in order to get rid of the burn in phase since there are no consecutive iterations to feed from. However, this is not a big issue since the prediction step only is performed once, not iteratively,  and the number of voxels of one CT image are smaller than that of all the voxels from all images used in the training set. For the prediction, $J = 1000$ was found to be sufficient. 

The important thing to note here is that, both for parameter estimation and CT prediction, the scaling in $N$ is linear which makes it feasible to fit the spatial models to large data sets such as multiple whole-brain images.

\section{An application to real data}
\label{sec:results}
Estimation of CT like images from MR data can be done in a great variety of ways, using for example atlases \citep{lit:arabi, lit:dowling}, or segmentation based \citep{lit:hsu} or combinations \citep{lit:siversson}. At present there are no way to directly compare these methods in terms of accuracy as they are based on different types of MR sequences acquired at different MR scanners with different field strength and with different coil solutions. The different studies report results for different areas of the body and have different inclusion and exclusion criteria's. To make a meaningful comparison of the proposed method we compare it to the previously published method of \citet{lit:adamCT} (GMM), using the same input data. 

This section presents the result from such a comparative cross-validation study. The models evaluated are:
\begin{enumerate}
\item Gaussian mixture model with spatially independence (GMM).
\item Gaussian mixture model with spatially dependence (GMMS). 
\item NIG mixture model with spatially independence (NIG).
\item NIG mixture model with spatially dependence (NIGS).
\end{enumerate}
Here the spatial dependence refer to the spatial prior of Section \ref{sec:spatialDependence}.

The data, described further in Section \ref{sec:data}, is three-dimensional images from scans of 14 patients. The results were analyzed using leave-one-out cross-validation with one fold for each patient. For each fold, parameters were estimated using the parameter estimation method described in Section \ref{sec:parameterEstimation} and data from all but but that fold. 
S-CT images was then generated for the fold using the estimated parameters and the MR images for the fold. Two s-CT images was generated for each model. One using the conditional expectation and one using the conditional median. The differences between the true CT images and the generated s-CT images were compared using MAE and RMSE. MAE is here chosen as the main metric for performance assessment since the amount of radiation that is attenuated on the way through a body is proportional to the accumulation of CT values over the radiation path. Hence, the conditional median should be the correct predictor to use for s-CT generation in order to assess the models \citep{lit:gneiting}.

The models ability to explain the distribution of the CT values conditioned on the MR images was also evaluated using the negatively oriented CRPS score \citep{lit:gneitingCRPS}, see \ref{app:CRPS}.

Choosing the parameter $K$, the number of classes, is not part of the estimation method. Hence, we are evaluating all the models for mixture classes ranging from 2 up to 10 in order to assess the sensitivity of this parameter.

Since neither the EM- nor the EM gradient- algorithm necessarily finds a global optima it is common to run the parameter estimation procedure several times with different and randomized initial values. 
In our implementation, the GMM was initialized using $15$ randomized starting values as well as two starting values acquired from estimating each class parameters from the data associated to it using a kmeans- and a hierarchical- clustering algorithm \citep{lit:Hastie}. 
For the other three models (GMMS, NIG and NIGS), only two initial values were used. One with the initial values derived from the estimated parameters of the GMM and one using just a set of constant values such as for instance a diagonal of ones as the precision matrices. 
Model selection among the initial values were chosen on the basis of lowest MAE of CT predictions on the training data. Performing model selection like this is better suited for our particular problem of s-CT generation compared to choosing the model with the highest likelihood. 

The main results can be seen in Section \ref{sec:mainresults} and some auxiliary results in Section \ref{sec:medianresults}.

\subsection{Details about data}
\label{sec:data}
The data used in this study consists of images from 14 different patients which were included in the study after oral and written consent. The study has been approved by the regional ethical review board of Ume\r{a} University. For each patient, one CT image and four MR images were acquired. For the MR images, two dual echo UTE sequences were used, one with a flip angle of $10^{\circ}$ and one with a flip angle of $30^{\circ}$. The UTE sequences sampled a first echo at $0.07$ ms and a second echo at $3.76$ ms. For both sequences the repetition time was $6$ ms. Two different flip angles and two different echo times give four possible combinations and hence four different MR images. 

The difference between images from the first and second echo indicate the presence of short T2* tissues. The differences between images acquired using the two flip angles indicate presence of T1 tissues. A short T2* is not only found in tissues with a short T2, but also in regions with rapid coherent dephasing, such as air-soft tissue and bone-soft tissue interfaces. Knowing the T1 information can help to separate these interfaces from T2. This is of interest since the T2 value is a good discriminator between bone, soft tissue and air \citep{lit:adamCT}.

\begin{figure}[t]
\centering

\begin{subfigure}{0.32\textwidth}
\centering
\includegraphics[width = 0.95\textwidth , keepaspectratio]{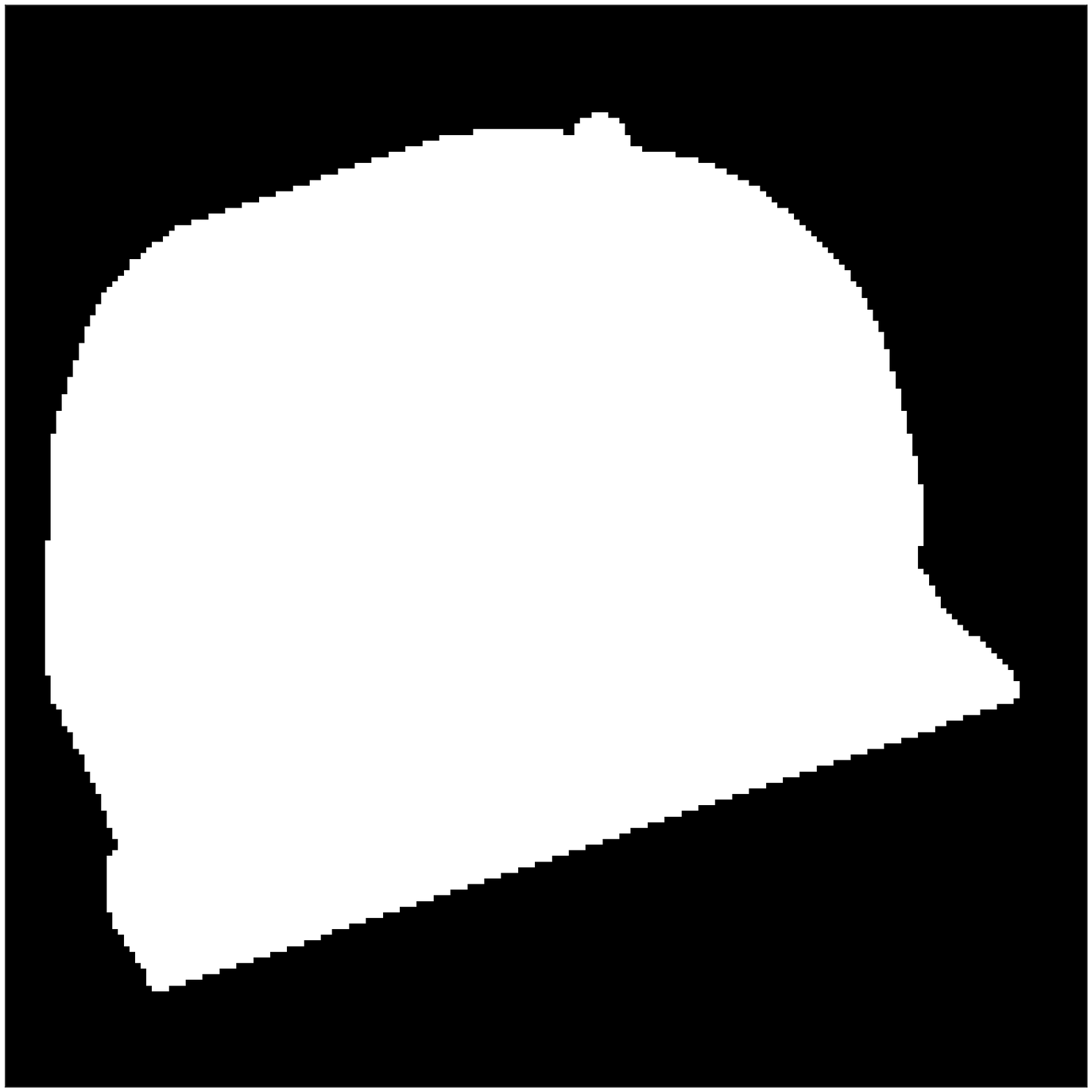} 
\caption{Binary mask}
\label{subfig:dataMask}
\end{subfigure}
\begin{subfigure}{0.32\textwidth}
\centering
\includegraphics[width = 0.95\textwidth , keepaspectratio]{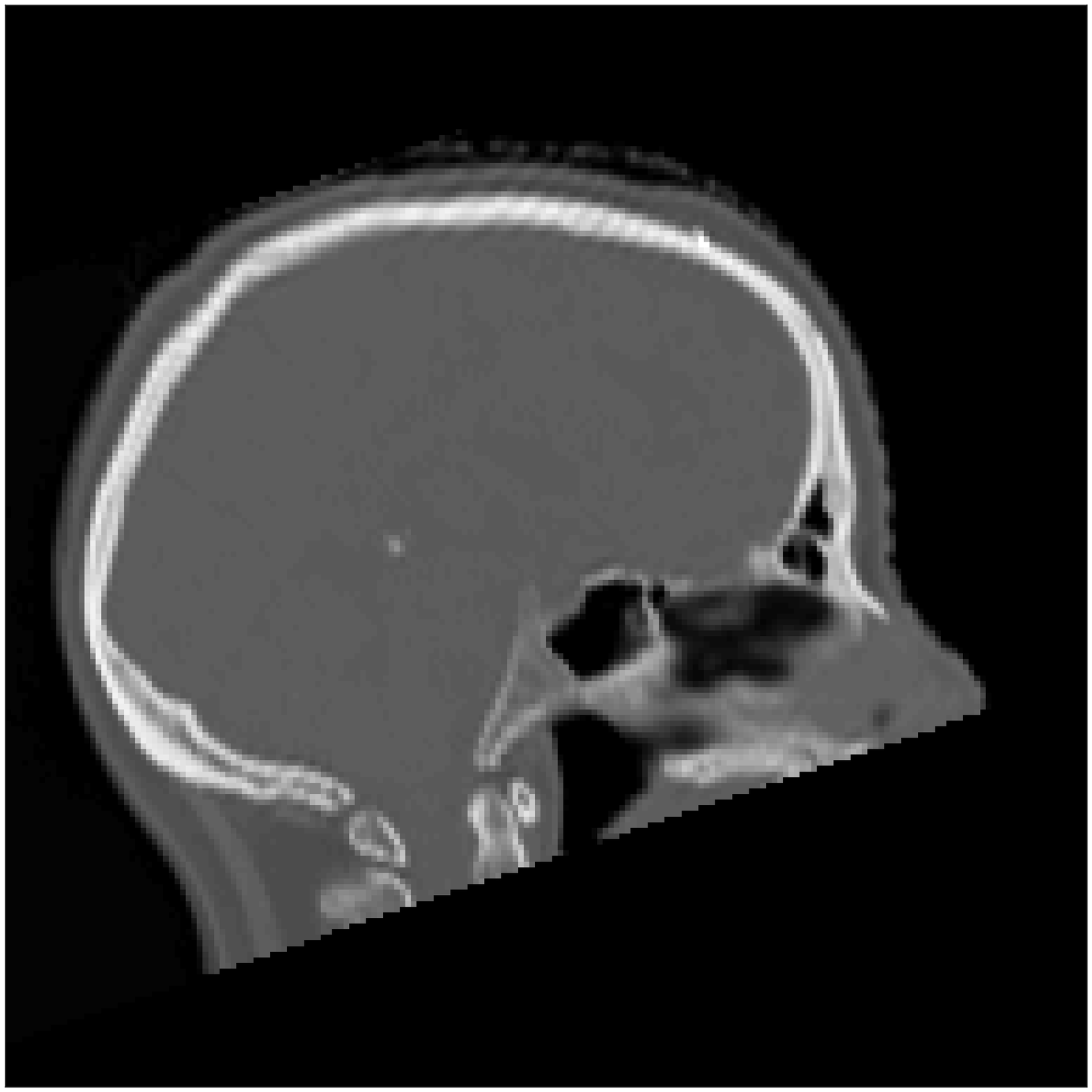} 
\caption{CT}
\label{subfig:dataCT}
\end{subfigure}
\begin{subfigure}{0.32\textwidth}
\centering
\includegraphics[width = 0.95\textwidth , keepaspectratio]{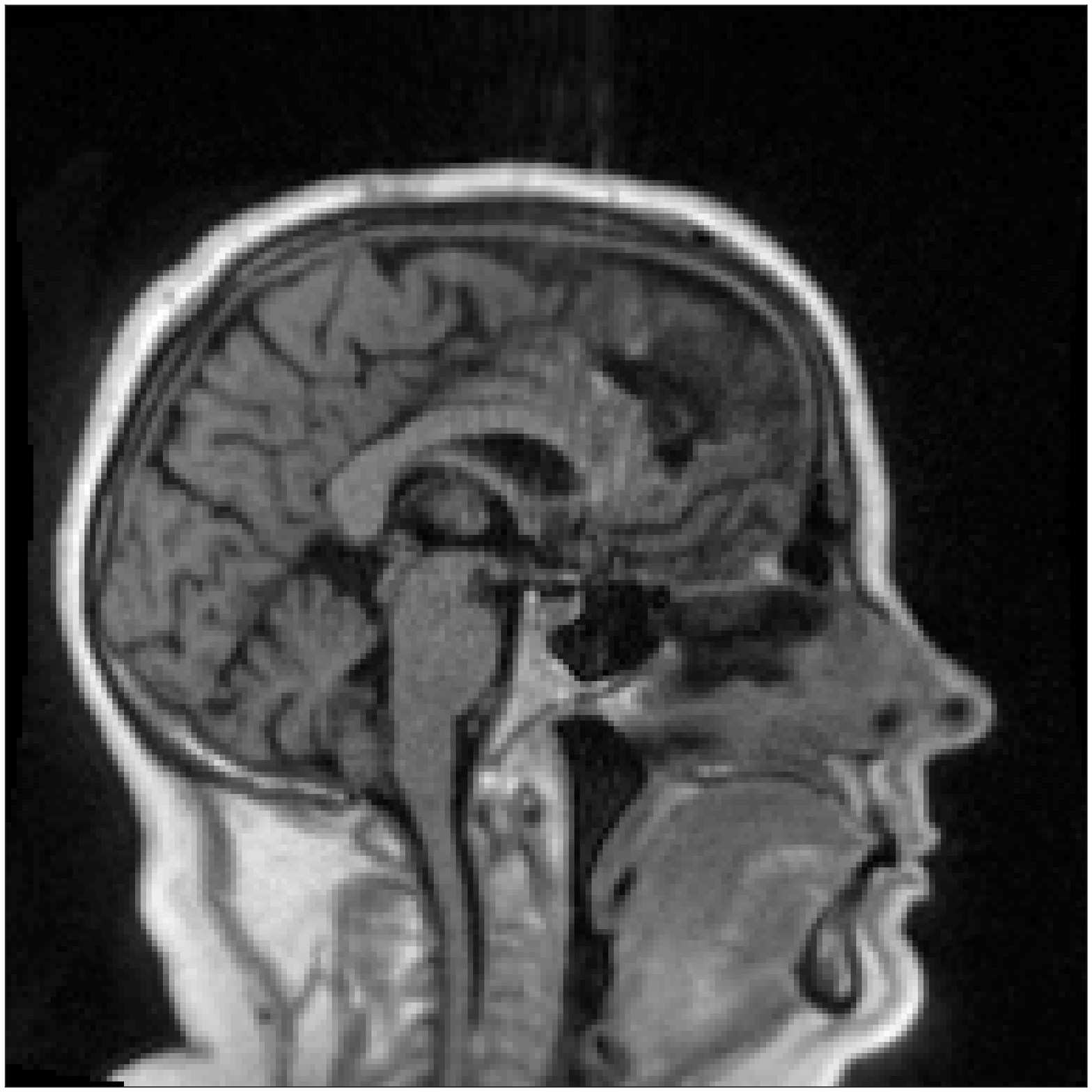}
\caption{First echo, $10^{\circ}$}
\label{subfig:dataMR1}
\end{subfigure}\\
\begin{subfigure}{0.32\textwidth}
\centering
\includegraphics[width = 0.95\textwidth , keepaspectratio]{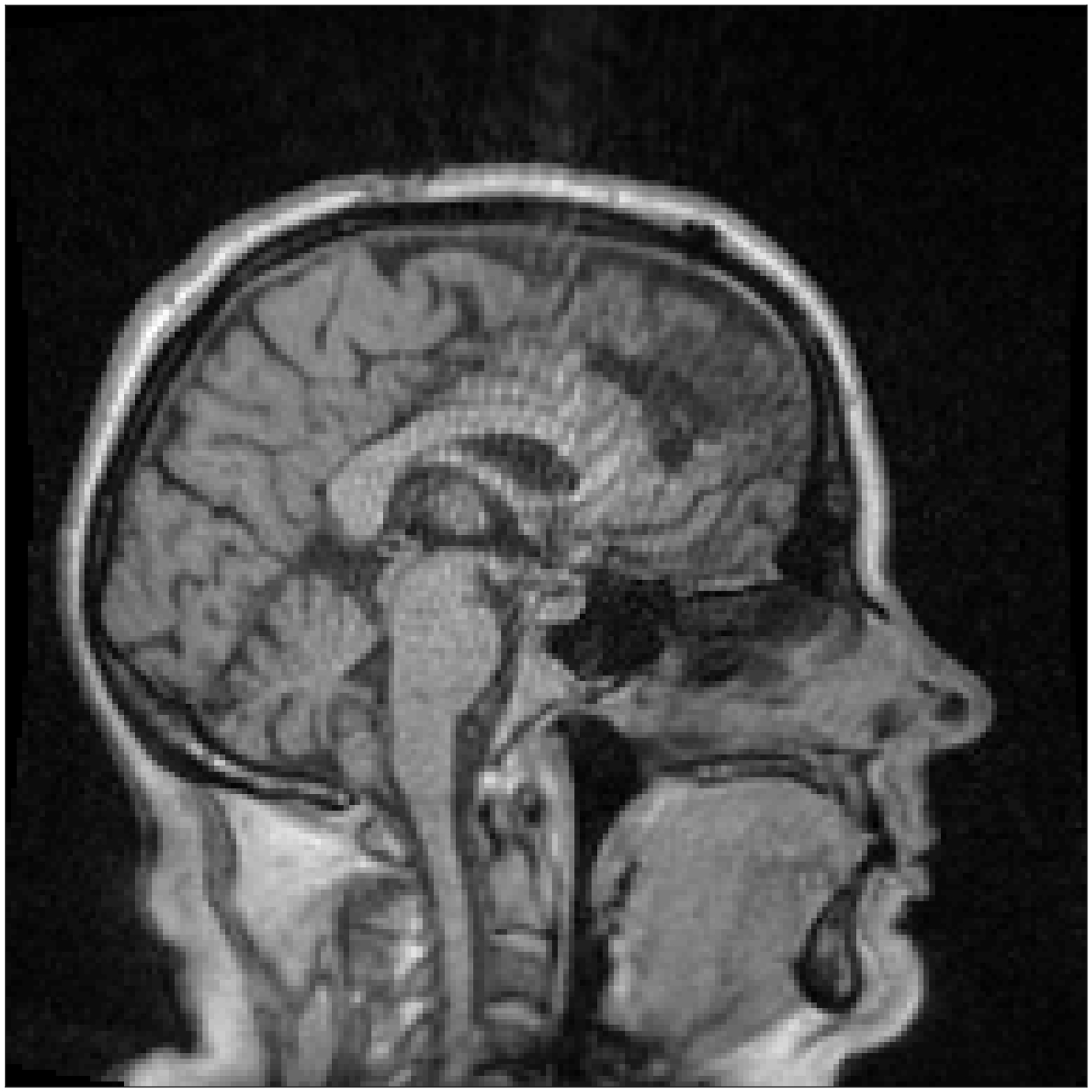} 
\caption{Second echo, $10^{\circ}$}
\label{subfig:dataMR2}
\end{subfigure}
\begin{subfigure}{0.32\textwidth}
\centering
\includegraphics[width = 0.95\textwidth , keepaspectratio]{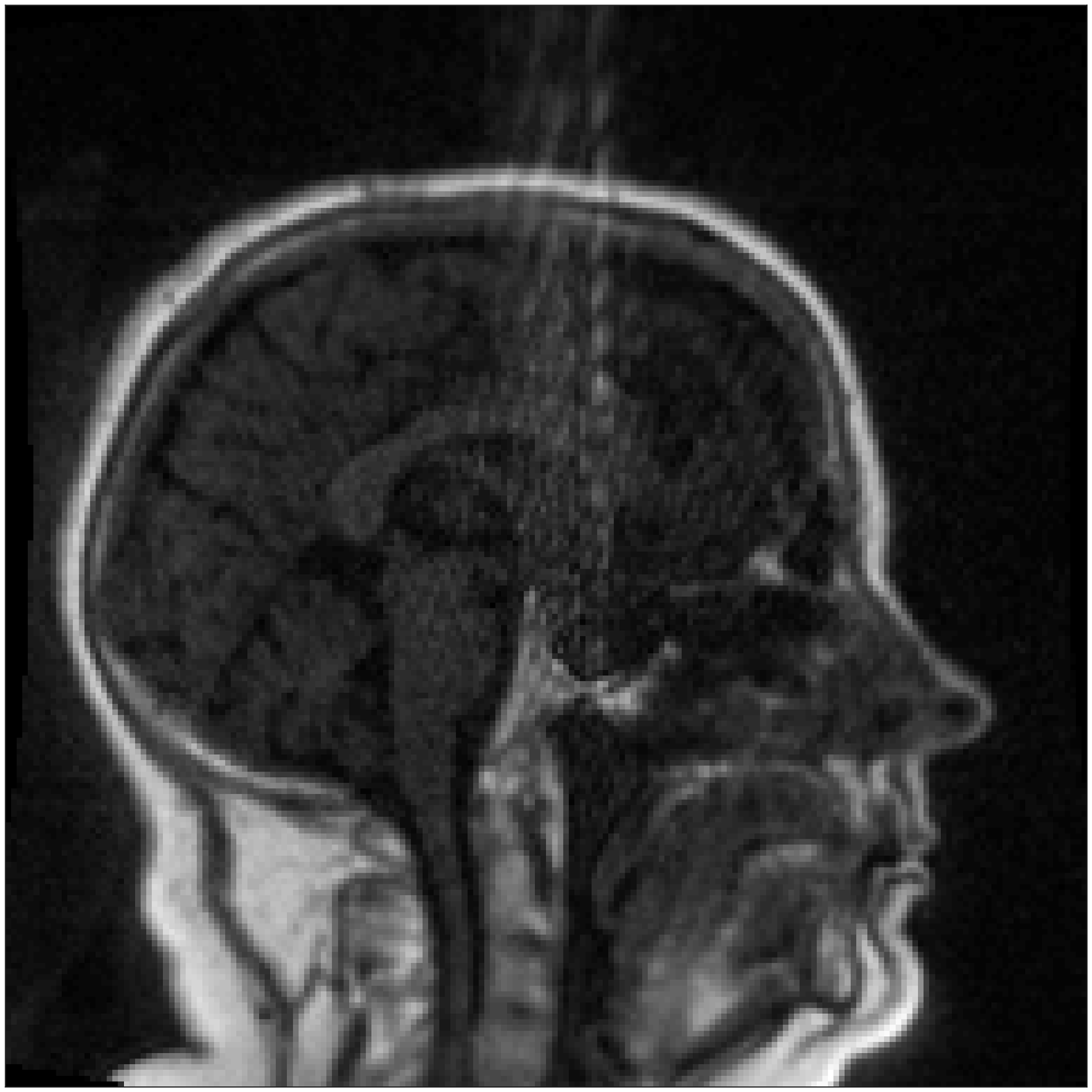}
\caption{First echo, $30^{\circ}$}
\label{subfig:dataMR3}
\end{subfigure}
\begin{subfigure}{0.32\textwidth}
\centering
\includegraphics[width = 0.95\textwidth , keepaspectratio]{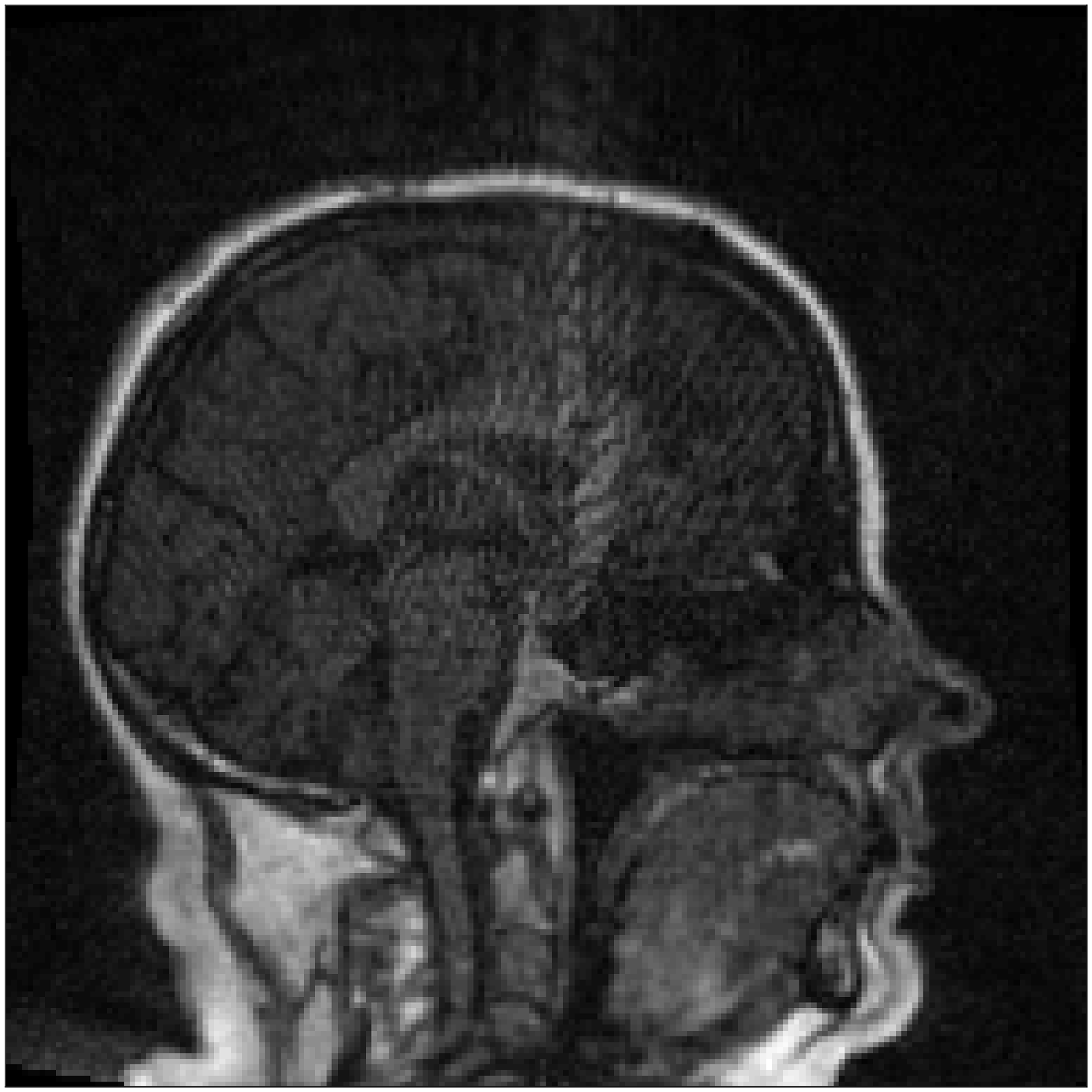}
\caption{Second echo, $30^{\circ}$}
\label{subfig:dataMR4}
\end{subfigure}
\caption{Binary data mask (panel a), CT image (panel b), The four MRI UTE sequences (panels c-f). }
\label{fig:data}
\end{figure}

All MR images were acquired with a 1.5 T Siemens Espree scanner. The UTE images were reconstructed to $192 \times 192 \times 192$ voxel bitmapped images with an isotropic resolution and a voxel size of 1.33 mm. The UTE sequences sampled the k-space radially with 30 000 radial spokes. CT images were acquired with a tube voltage of between 120 kV and 130 kV on either a GE Lightspeed Plus, Siemens Emotion 6 or GE Discovery 690. The in-plane pixel size varied between 0.48 mm to 1.36 mm and the slice thickness between 2.5 mm and 3.75 mm.  Images of the same patient were co-registered and resampled to achieve voxel-wise correspondence between all five modes. A binary mask excluding most of the air surrounding the head was computed from the images and used to remove unnecessary data. Furthermore, to reduce the execution time of the parameter estimation phase, only  11 slices in the middle of the head of each patient was used during the parameter estimation phase, but all slices were used during the prediction phase (s-CT generation). Additional details concerning the data can be found in \citet{lit:adamCT}. Data from one slice of a patient is shown in Figure \ref{fig:data}.

\begin{figure}[t]
\centering
\includegraphics[width = 0.7\textwidth, keepaspectratio]{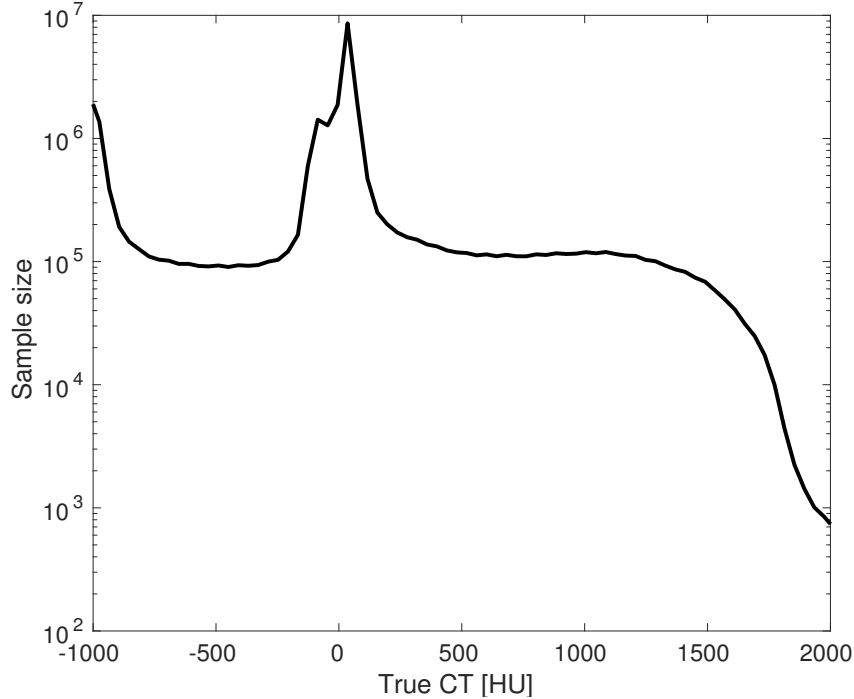}
\caption{Smoothed histogram of CT values in the total data set of all 14 patients. Note the logarithmic scale on the y-axis.  }
\label{fig:samplesVSCT}
\end{figure}

Figure \ref{fig:samplesVSCT} shows a smoothed histogram of the CT values for all patients. Note that most voxels have CT values close to zero HU. This corresponds to soft tissue which makes up the main volume of the head. The peak at around -1000 HU in the histogram corresponds to air. The presence of air is partly due to the cavities in the nasal region, sinuses and throat but also partly due to that the binary mask around the head is not completely tight and allow air in between the actual outline of the head and what the mask cuts away. The higher CT values (typically around 600 to 1500) corresponds to bone. 

The very high CT values ($> 1500$) correspond to streak artifacts or interpolation errors in the resampling procedure.

\subsection{Main results}
\label{sec:mainresults}
\begin{figure}[t]
\centering
\begin{subfigure}{0.47 \linewidth}
\centering
\includegraphics[width = \textwidth, keepaspectratio]{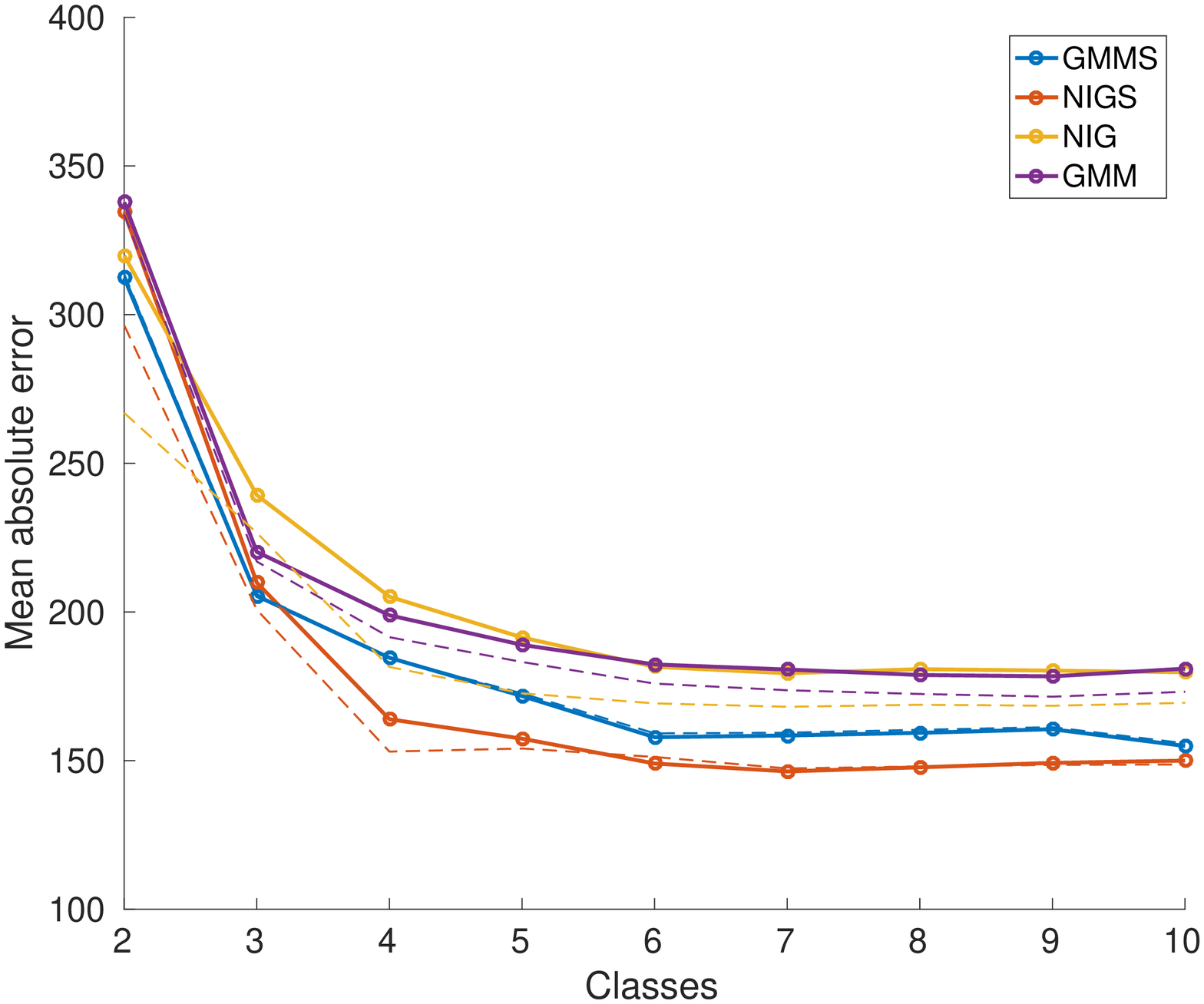}
\caption{MAE}
\label{fig:maeErrors}
\end{subfigure} 
\begin{subfigure}{0.47 \linewidth}
\centering
\includegraphics[width = \textwidth, keepaspectratio]{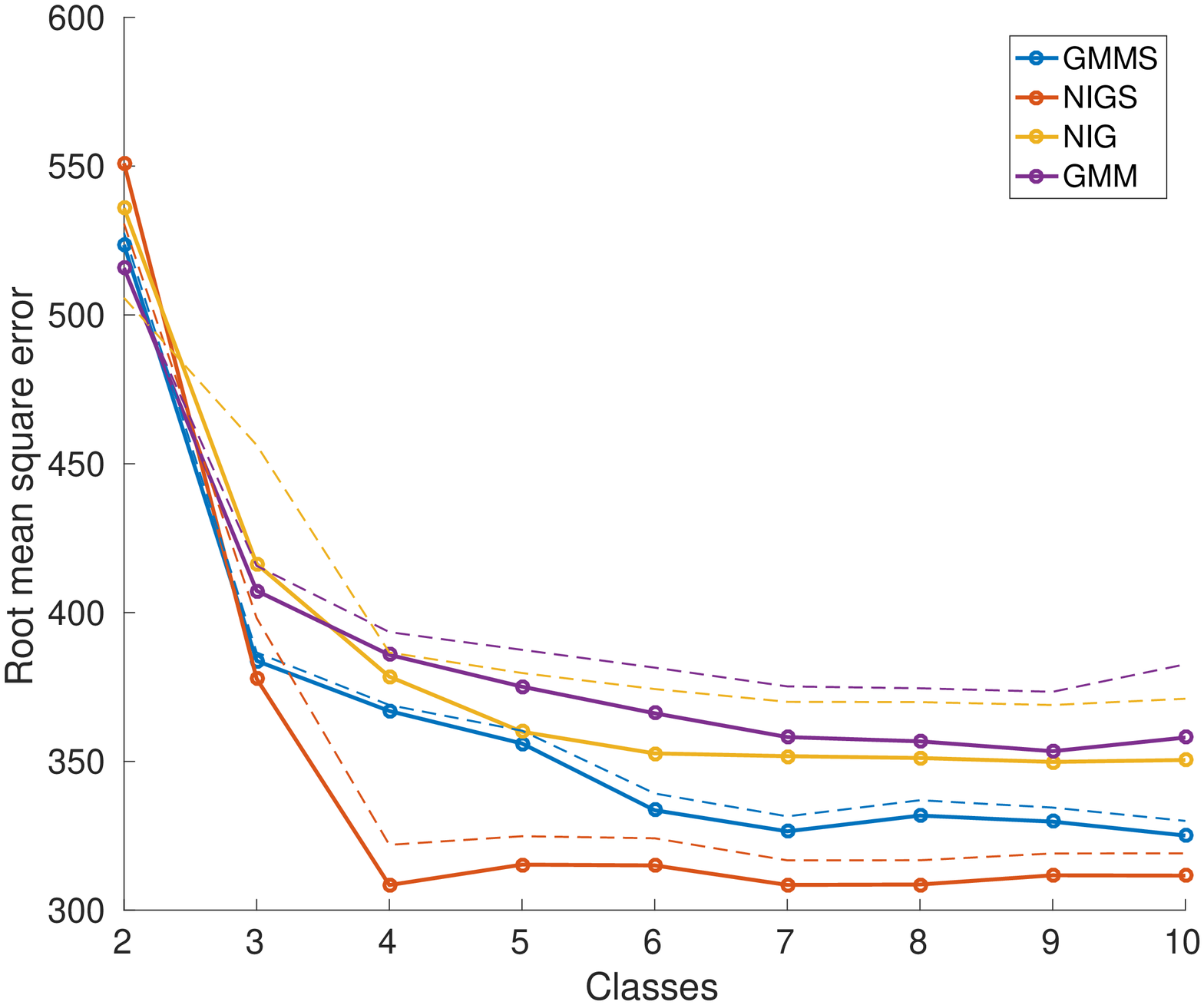}
\caption{RMSE}
\label{fig:rmseErrors}
\end{subfigure} \\
\caption{Errors of s-CT images compared to the number of classes in the mixture model. Errors shown for both conditional mean (thick solid lines) and conditional median predictions (thin dashed lines).  }
\label{fig:Errors}
\end{figure}

The cross-validation study was carried out for each of the four models described above, and each model was tested with the number of mixture classes ranging from two up to ten. MAE and RMSE values from the study can be seen in Figure \ref{fig:Errors} and in Table \ref{table:optimMAE}. The lowest MAE value ($146.4$ HU) was attained using the NIGS model with seven classes and conditional mean. Except for two- or three- class models the NIGS model followed by the GMMS had the lowest prediction error both in MAE and RMSE. 

The conditional median improves the MAE for the non-spatial models but do not affect the spatial ones as clearly. For the GMMS there is almost no difference between using mean or median as predictor. For the NIGS it improves MAE when using two to five classes but performs similarly when $K > 5$. For RMSE it is a consistent drawback of using median instead of mean as the predictor, as expected.

\begin{table}[t]
\centering
\caption{Prediction errors of the models at the number of classes were MAE reached its minimum for each model. The models are compared both using mean and median as predictor. "Ratios" compare the corresponding error with the reference model (GMM with mean predictor). }
\begin{tabular}{l | l | c | c c | c c}
Model & Predictor & Classes & MAE & MAE Ratio & RMSE & RMSE Ratio \\
\hline
GMM & Mean & 9 & 178.3 & 100\% & 353.4 & 100\% \\
GMM & Median & 9 & 171.5 & 96.2\% & 373.4 & 105.7\% \\
GMMS & Mean & 10 & 154.9 & 86.9\% & 325.1 & 92.0\% \\
GMMS & Median& 10 & 155.8 & 87.4\% & 330.0 & 93.4\% \\
NIG & Mean & 7 & 179.4 & 100.5\% & 351.7 & 99.5\% \\
NIG & Median& 7 & 168.1 & 94.3\% & 370.0 & 104.7\% \\
NIGS & Mean & 7 & 146.4 & 82.1\% & 308.5 & 87.3\% \\
NIGS & Median& 7 & 147.3 & 82.6\% & 316.7 & 89.6\% \\
\hline
\end{tabular}
\label{table:optimMAE}
\end{table}

\begin{figure}[t]
\centering
\begin{subfigure}{0.45 \linewidth}
\centering
\includegraphics[width = 0.95\textwidth, keepaspectratio]{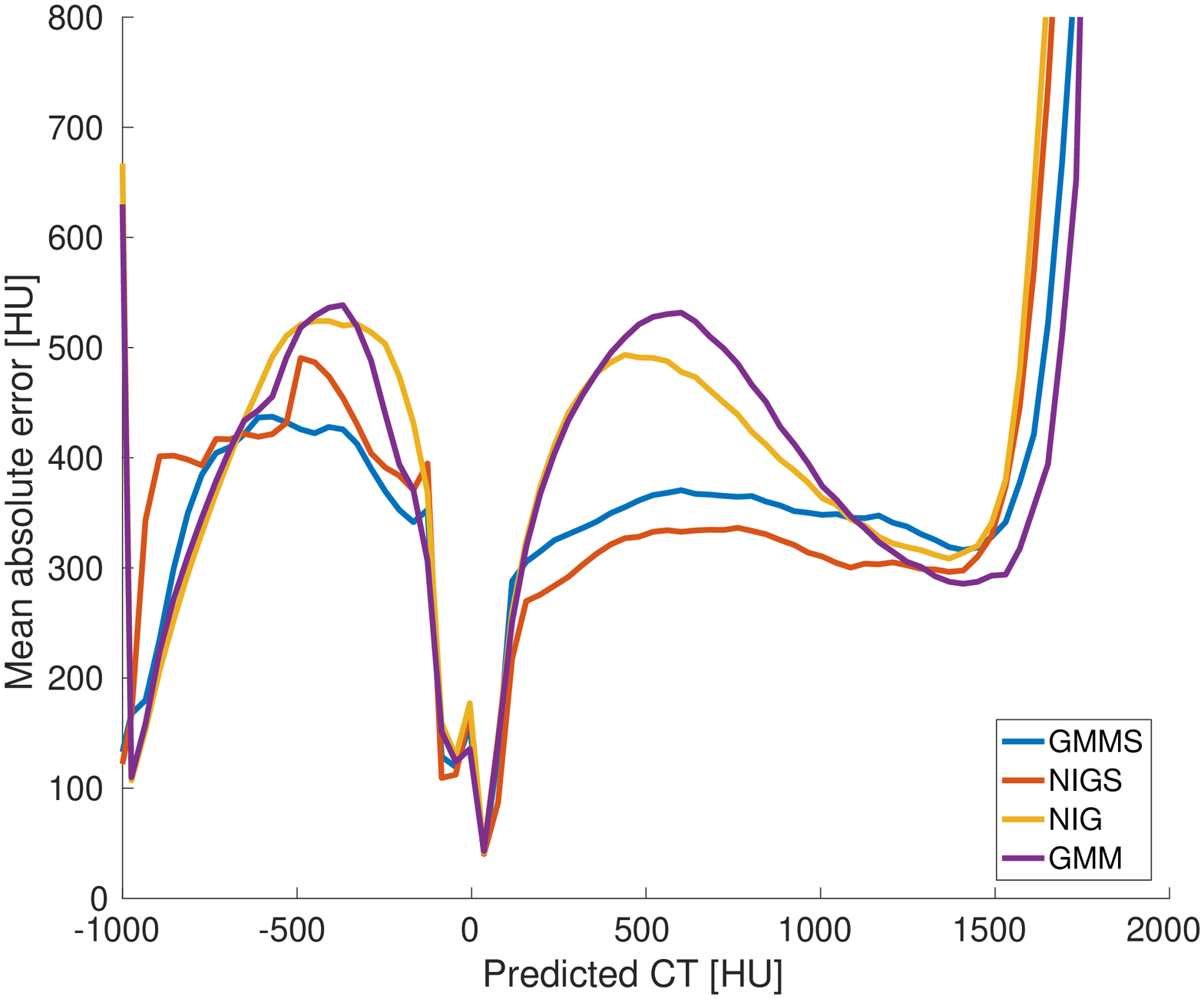}
\caption{Mean absolute errors. }
\label{subfig:maeVSCT}
\end{subfigure}
\begin{subfigure}{0.45 \linewidth}
\centering
\includegraphics[width = 0.95\textwidth, keepaspectratio]{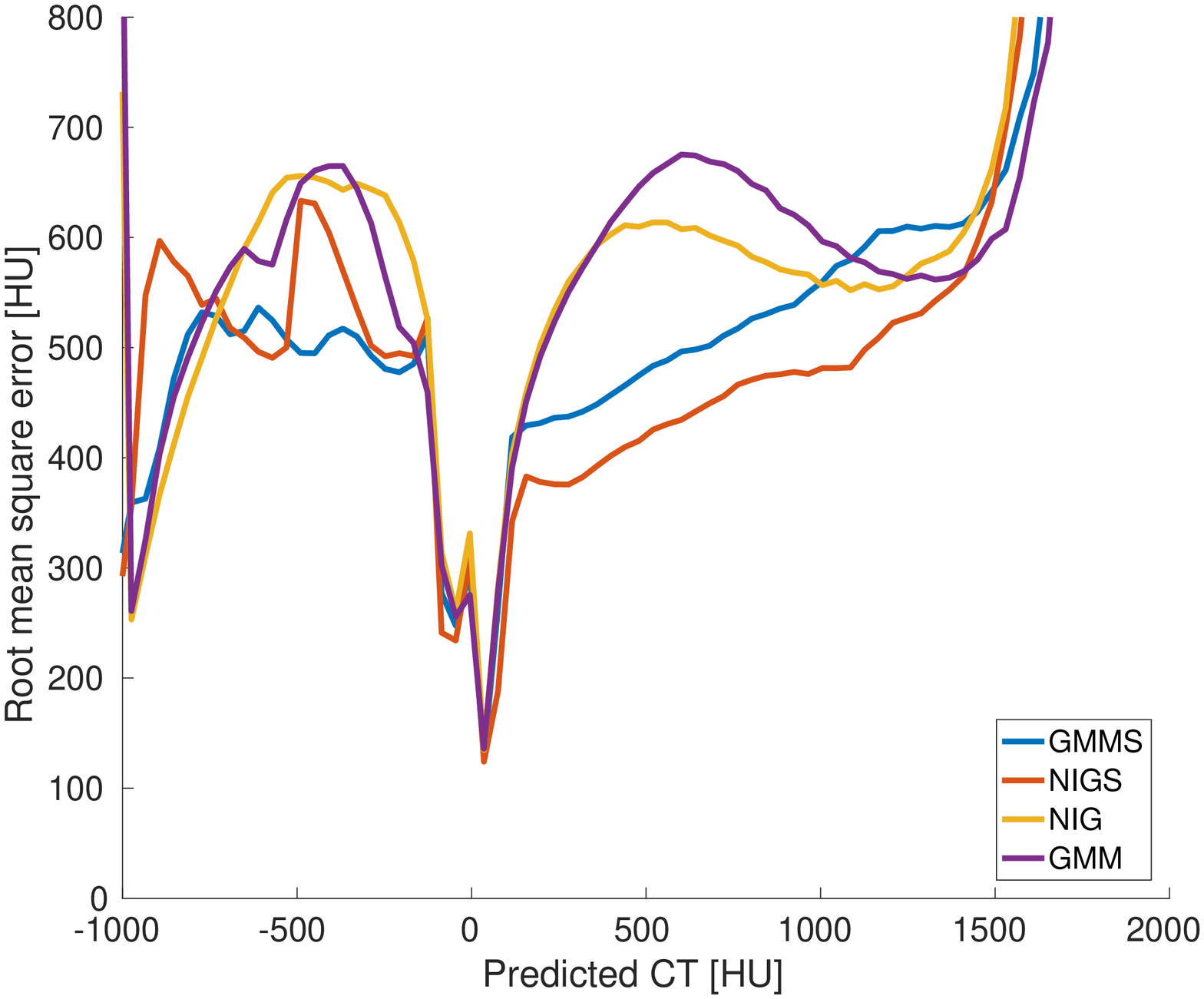}
\caption{Root mean square error.}
\label{subfig:rmseVSCT}
\end{subfigure} \\
\begin{subfigure}{0.45 \linewidth}
\centering
\includegraphics[width = 0.95\textwidth, keepaspectratio]{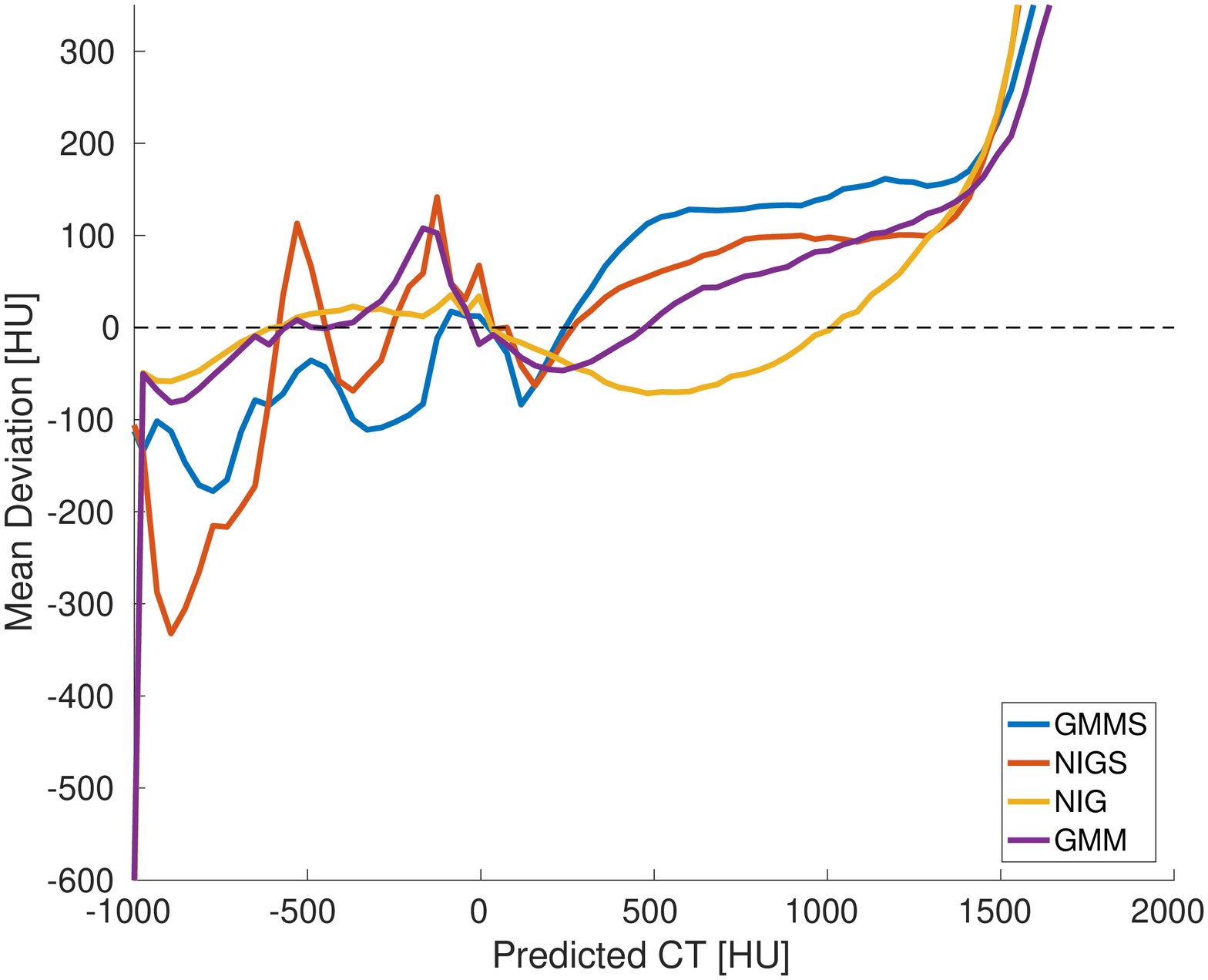}
\caption{Mean error.}
\label{subfig:devVSCT}
\end{subfigure}
\begin{subfigure}{0.45 \linewidth}
\centering
\includegraphics[width = 0.95\textwidth, keepaspectratio]{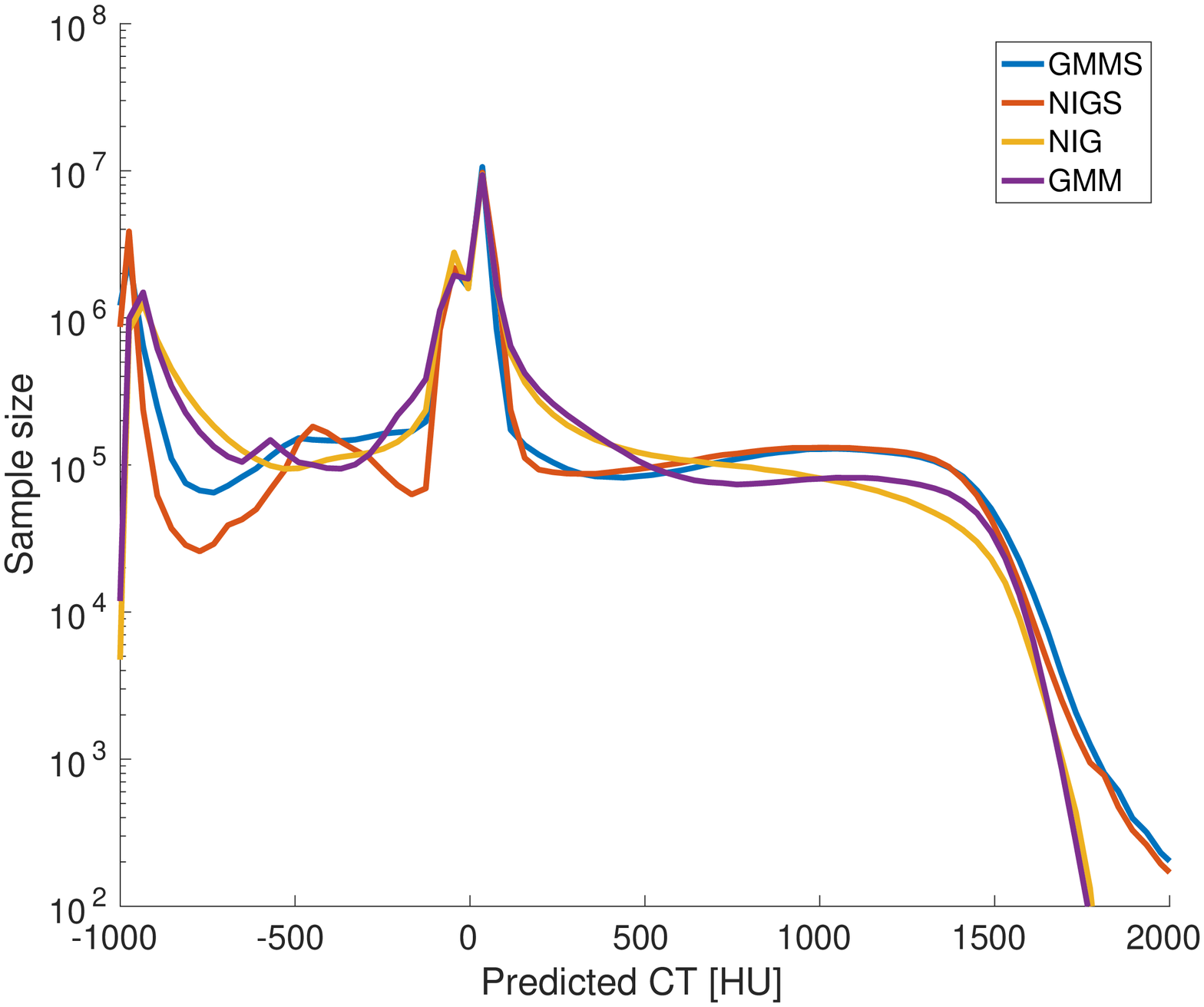}
\caption{Sample size density. }
\label{subfig:samplesVSCT}
\end{subfigure}
\caption{The MAE, RMSE, mean error and sample size density as functions of the predicted CT values [HU] for the four different models at the number of classes that minimized MAE. }
\label{fig:VSCT}
\end{figure}

Figure \ref{fig:VSCT} shows prediction errors as functions of predicted CT values. Comparing the panels one can note that the prediction errors for all models are smaller for voxels where the predicted CT value is in an interval with more frequent CT values. Here this corresponds to soft tissue (around $0$ HU) and air (around $-1000$ HU).

Figure \ref{subfig:devVSCT} shows the bias of the predictions. Here, the spatial models seem to have a generally bigger bias. However, at the same time the spatial models have a generally lower MAE and RMSE, especially when predicting CT values above $0$, i.e. bone. This suggests that the variance of the estimates are smaller for the spatial models.

\begin{figure}[t]
\centering
\includegraphics[width = 0.45\textwidth, keepaspectratio]{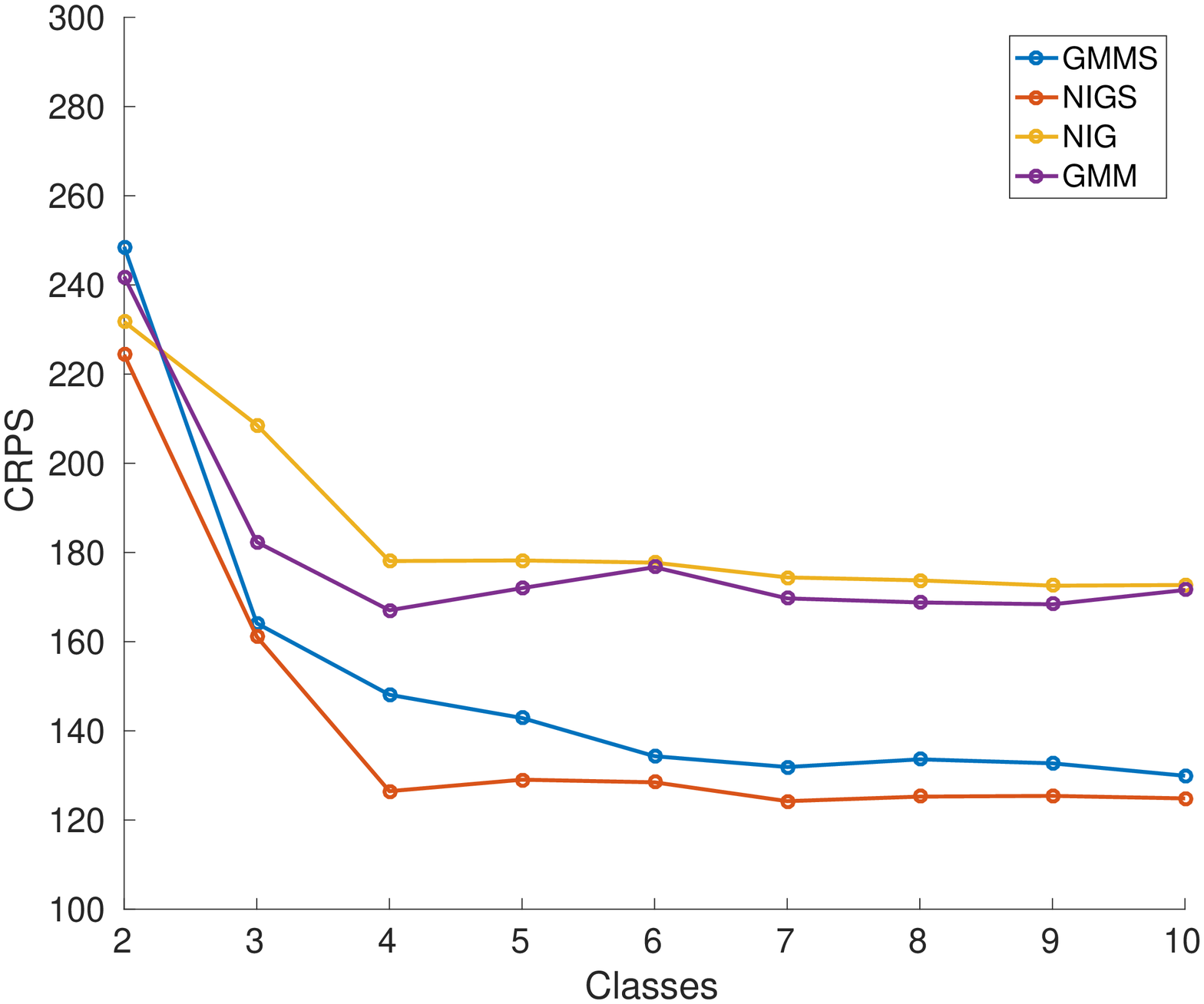}
\caption{CRPS* for all models and number of classes.}
\label{fig:crps}
\end{figure}

The negatively oriented CRPS (from now on referred to as the CRPS*) is a measure of how well a probability distribution explains the observed data, see \ref{app:CRPS} for further details. A small CRPS* indicates a good distributional fit. Note that, CRPS* is only associated with the conditional model and not with the chosen prediction function derived from it. 
Figure \ref{fig:crps} show the mean CRPS* over all predicted voxels for each model and number of classes. As can be seen, the CRPS follow the same behavior as the errors of Figures \ref{fig:Errors} except that the NIG model performs worse than the GMM.

\begin{figure}[t]
\centering

\begin{tabular}{rcccccc}

&CT & & Error & & Std \\
\vspace*{0.2cm}
True CT & 
\begin{subfigure}{0.16\linewidth}
\centering
\includegraphics[width = 0.99\textwidth, keepaspectratio]{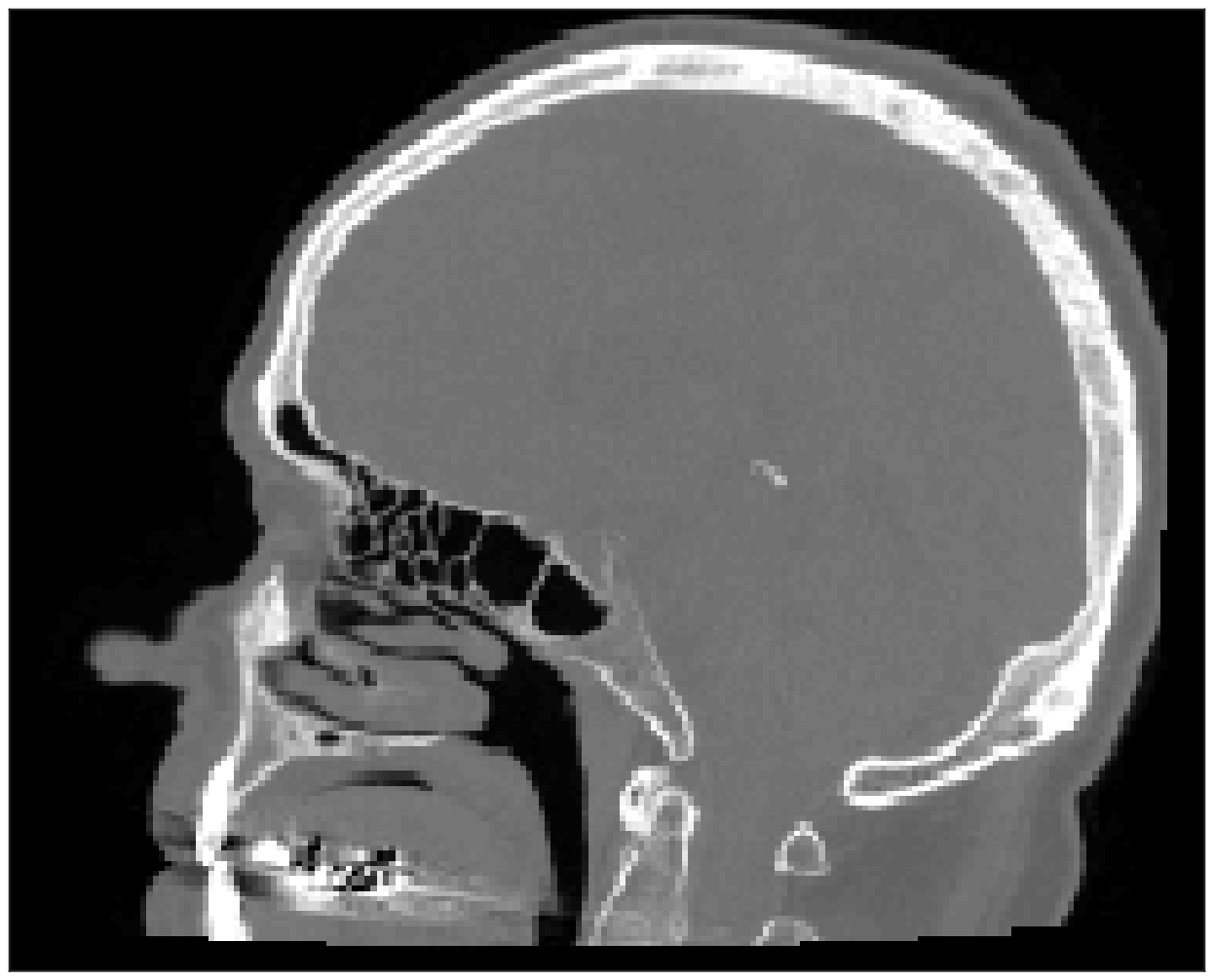}
\end{subfigure} 
\\

\vspace*{0.2cm}
GMM & 
\begin{subfigure}{0.16\linewidth}
\centering
\includegraphics[width = 0.99\textwidth, keepaspectratio]{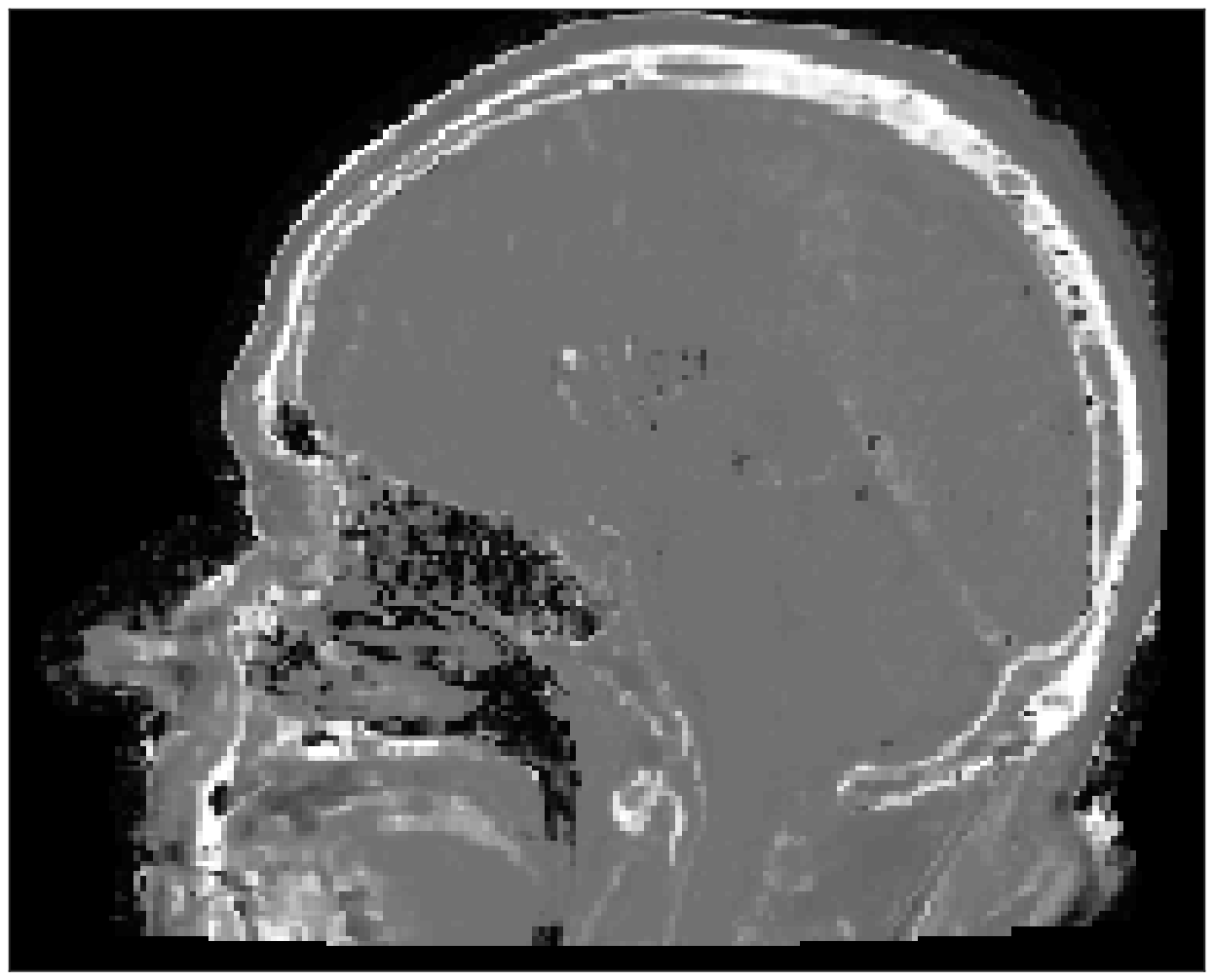}
\end{subfigure} &
\multirow{4}{*}{ 
\begin{subfigure}{0.05\textwidth} \includegraphics[width = \textwidth, height = 5.7 cm ]{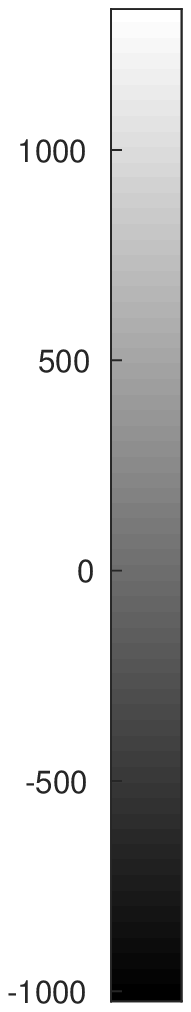}
\end{subfigure} 
} &

\begin{subfigure}{0.16\linewidth}
\centering
\includegraphics[width = 0.99\textwidth, keepaspectratio]{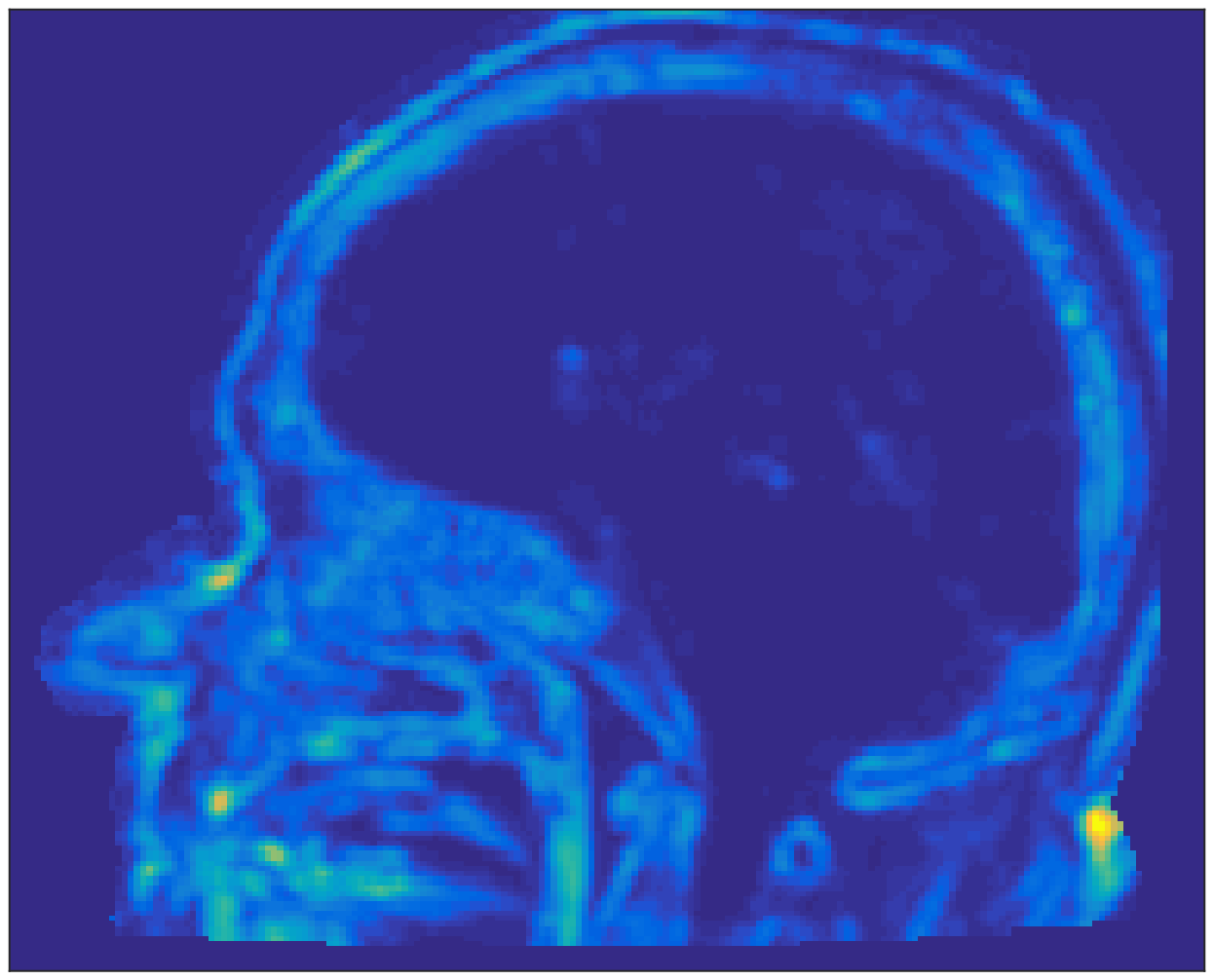}
\end{subfigure} &
\multirow{4}{*}{ 
\begin{subfigure}{0.05\textwidth} \includegraphics[width = \textwidth, height = 5.7 cm ]{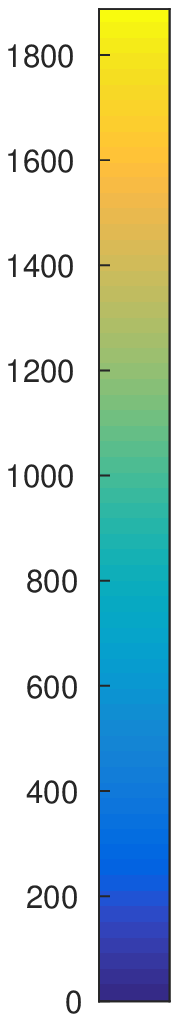}
\end{subfigure} 
} &
\begin{subfigure}{0.16\linewidth}
\centering
\includegraphics[width = 0.99\textwidth, keepaspectratio]{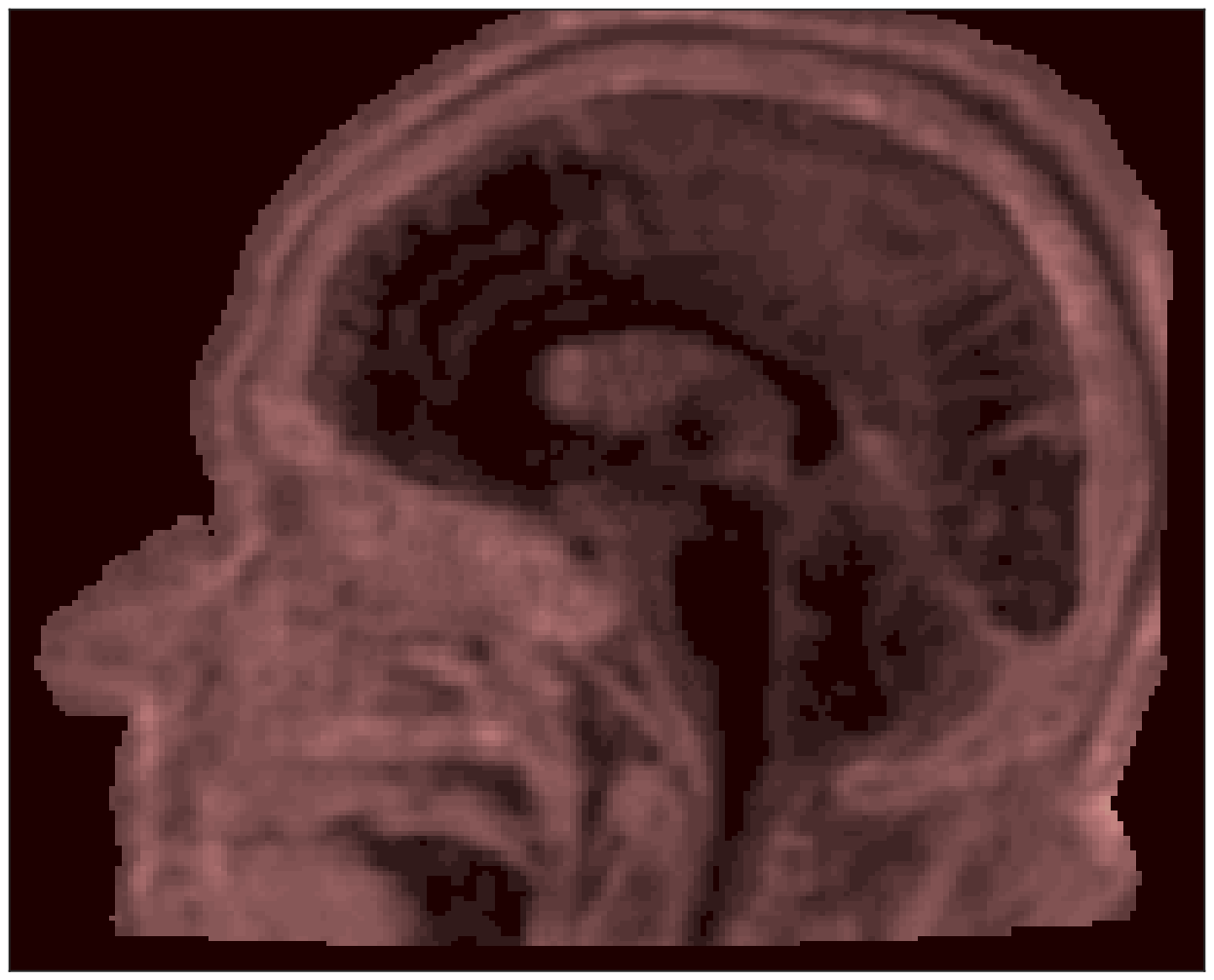}
\end{subfigure} &
\multirow{4}{*}{ 
\begin{subfigure}{0.05\textwidth} \includegraphics[width = \textwidth , height = 5.7 cm ]{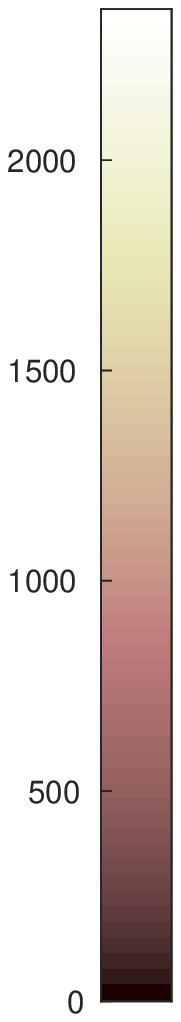}
\end{subfigure} 
} 
\\

\vspace*{0.2cm}
GMMS & 
\begin{subfigure}{0.16\linewidth}
\centering
\includegraphics[width = 0.99\textwidth, keepaspectratio]{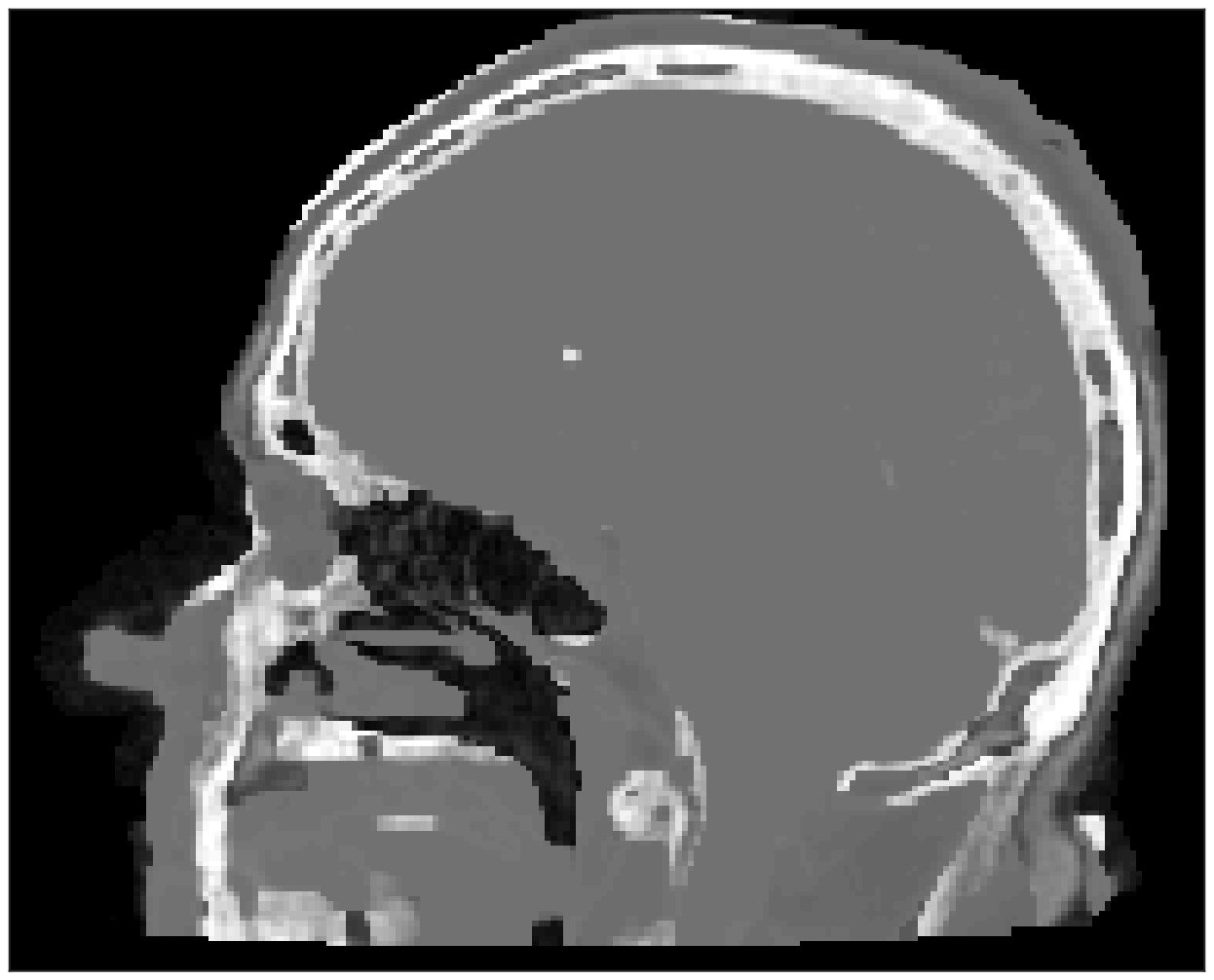}
\end{subfigure} & &
\begin{subfigure}{0.16\linewidth}
\centering
\includegraphics[width = 0.99\textwidth, keepaspectratio]{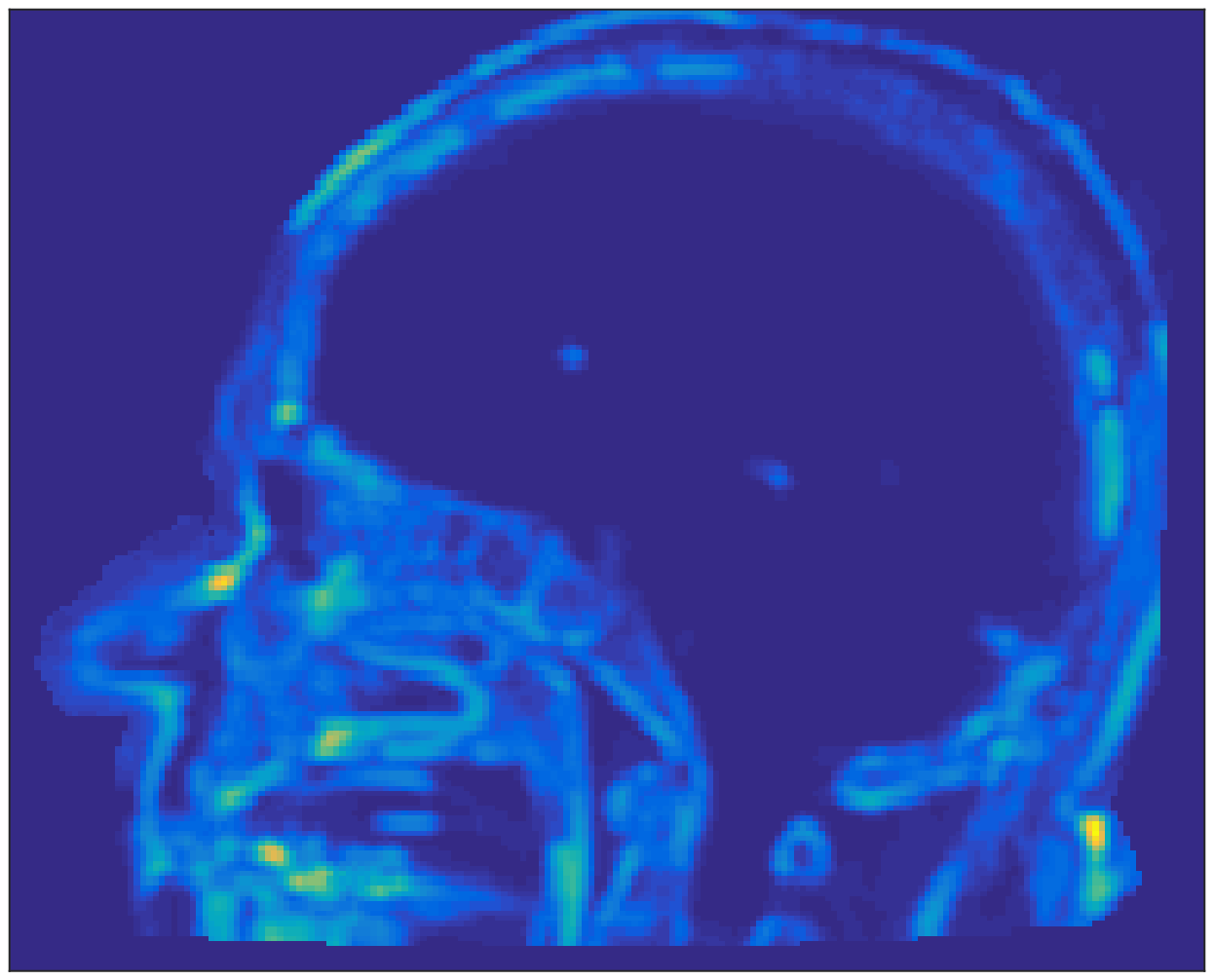}
\end{subfigure} &&
\begin{subfigure}{0.16\linewidth}
\centering
\includegraphics[width = 0.99\textwidth, keepaspectratio]{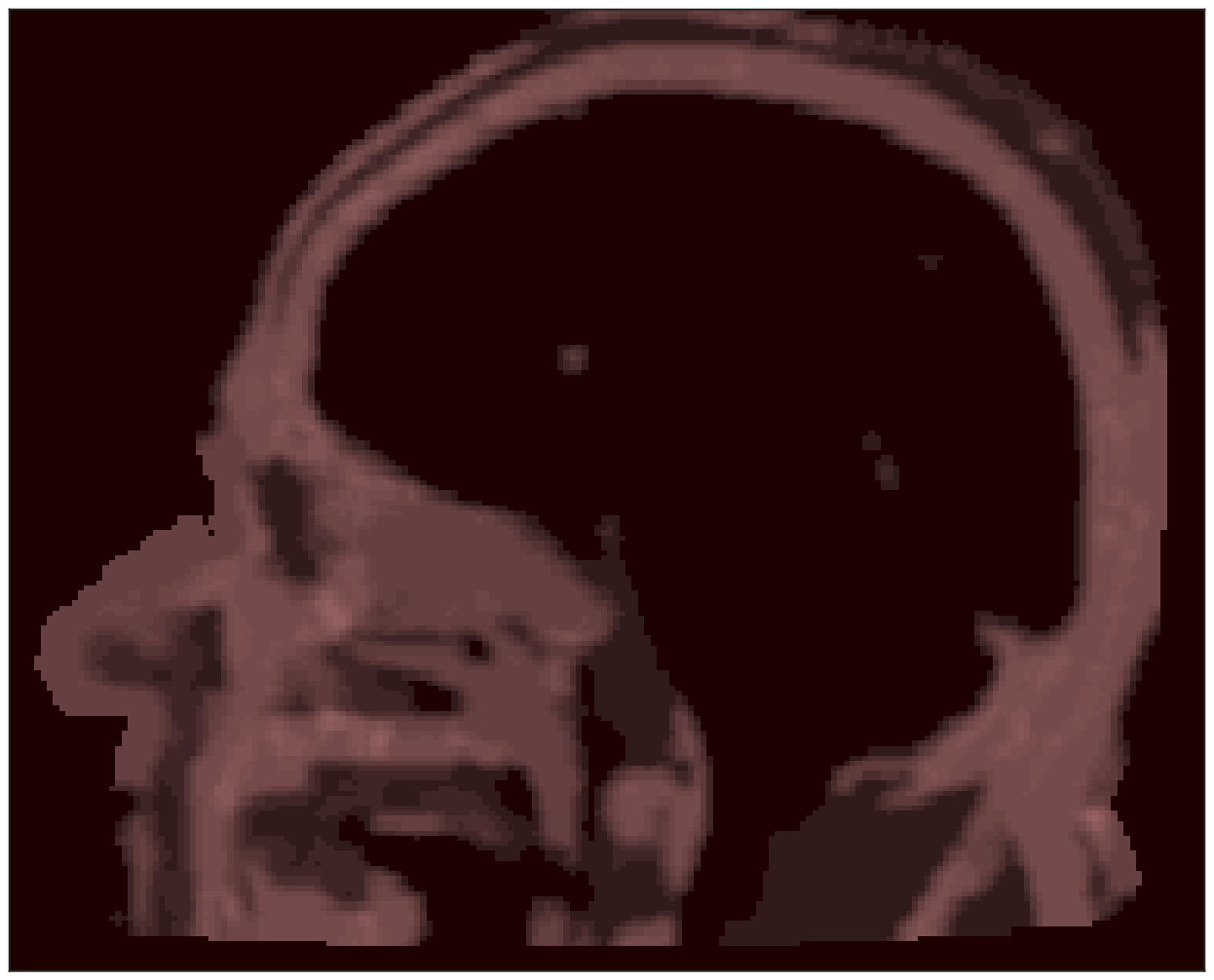}
\end{subfigure}
\\

\vspace*{0.2cm}
NIG & 
\begin{subfigure}{0.16\linewidth}
\centering
\includegraphics[width = 0.99\textwidth, keepaspectratio]{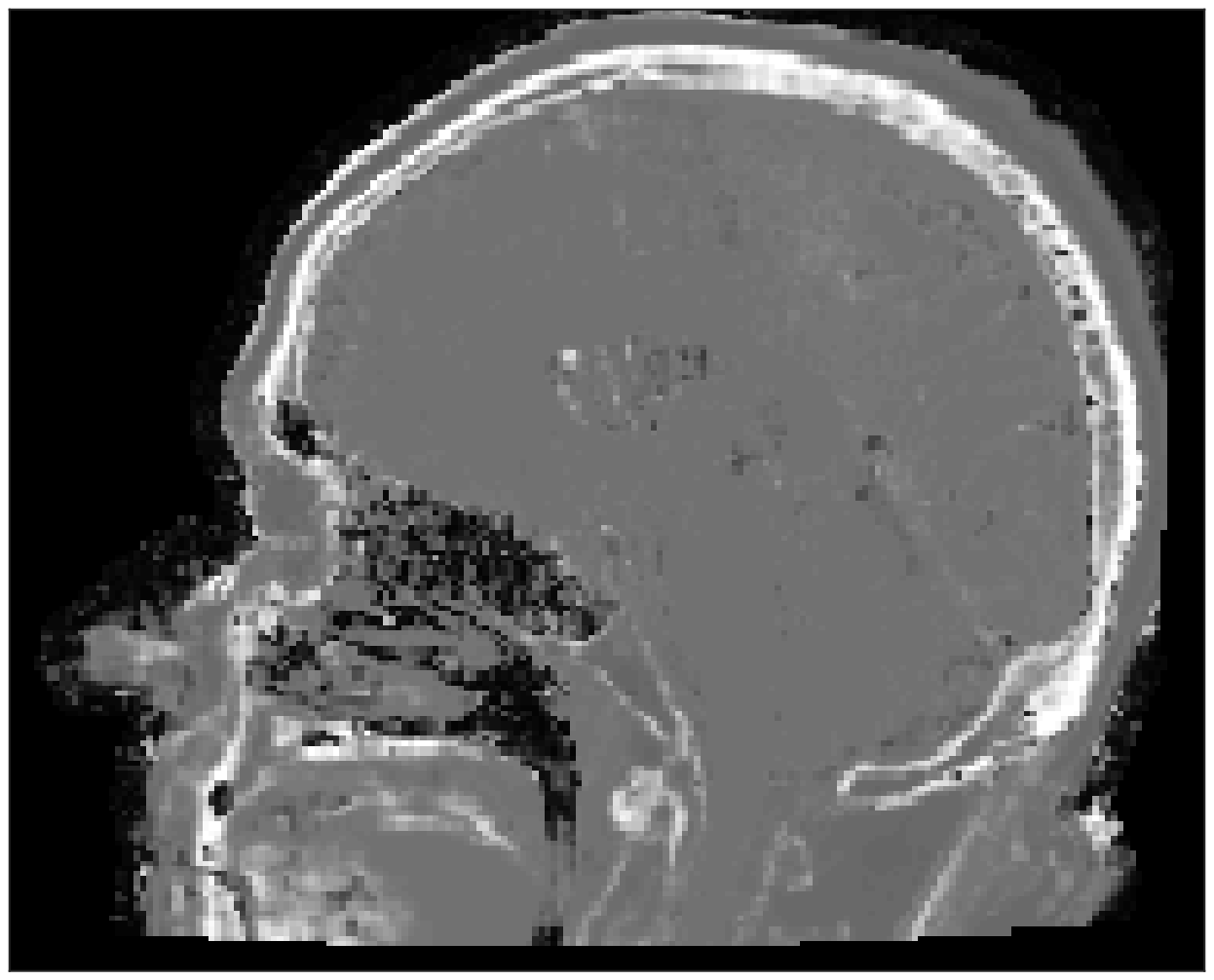}
\end{subfigure} & &
\begin{subfigure}{0.16\linewidth}
\centering
\includegraphics[width = 0.99\textwidth, keepaspectratio]{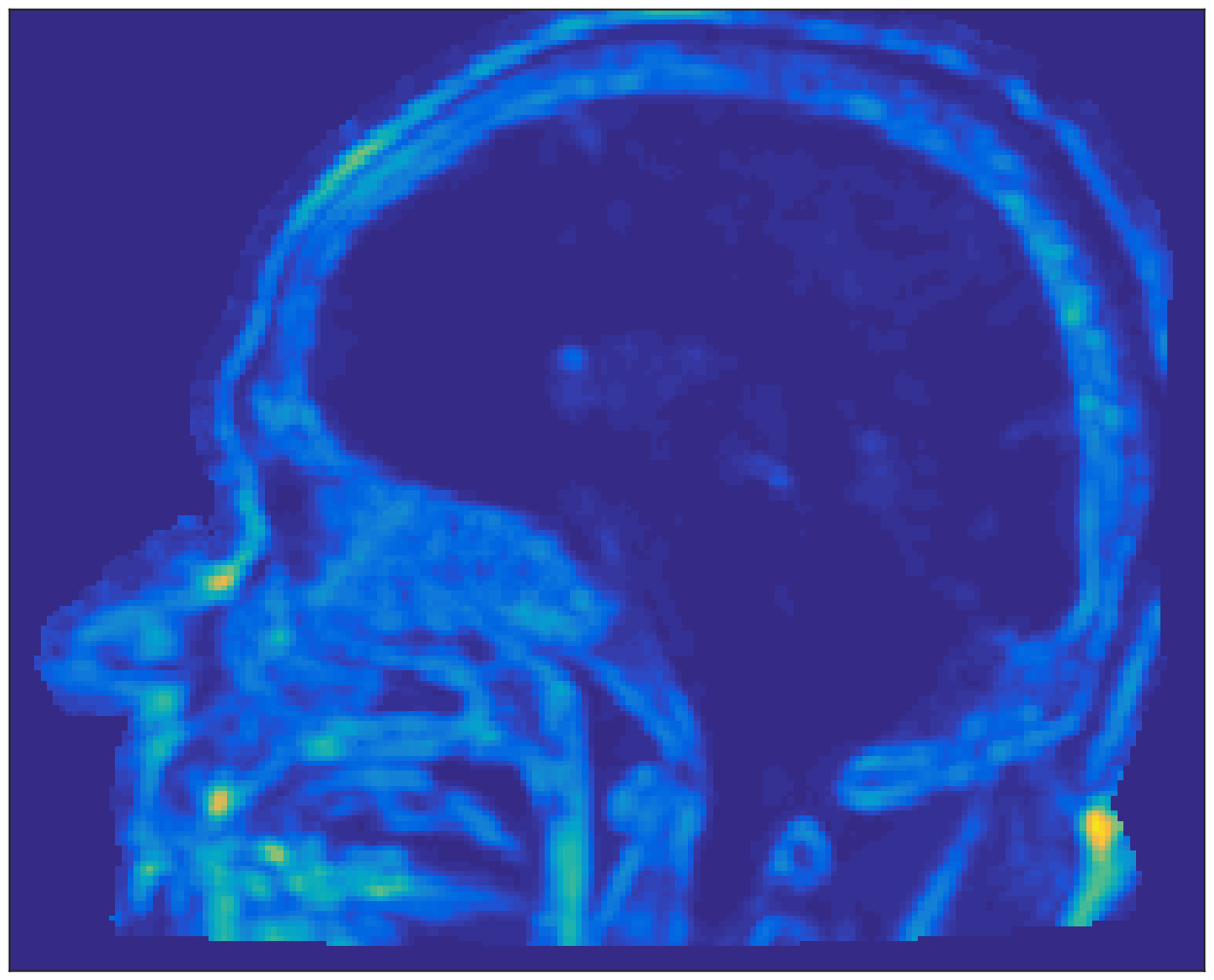}
\end{subfigure} &&
\begin{subfigure}{0.16\linewidth}
\centering
\includegraphics[width = 0.99\textwidth, keepaspectratio]{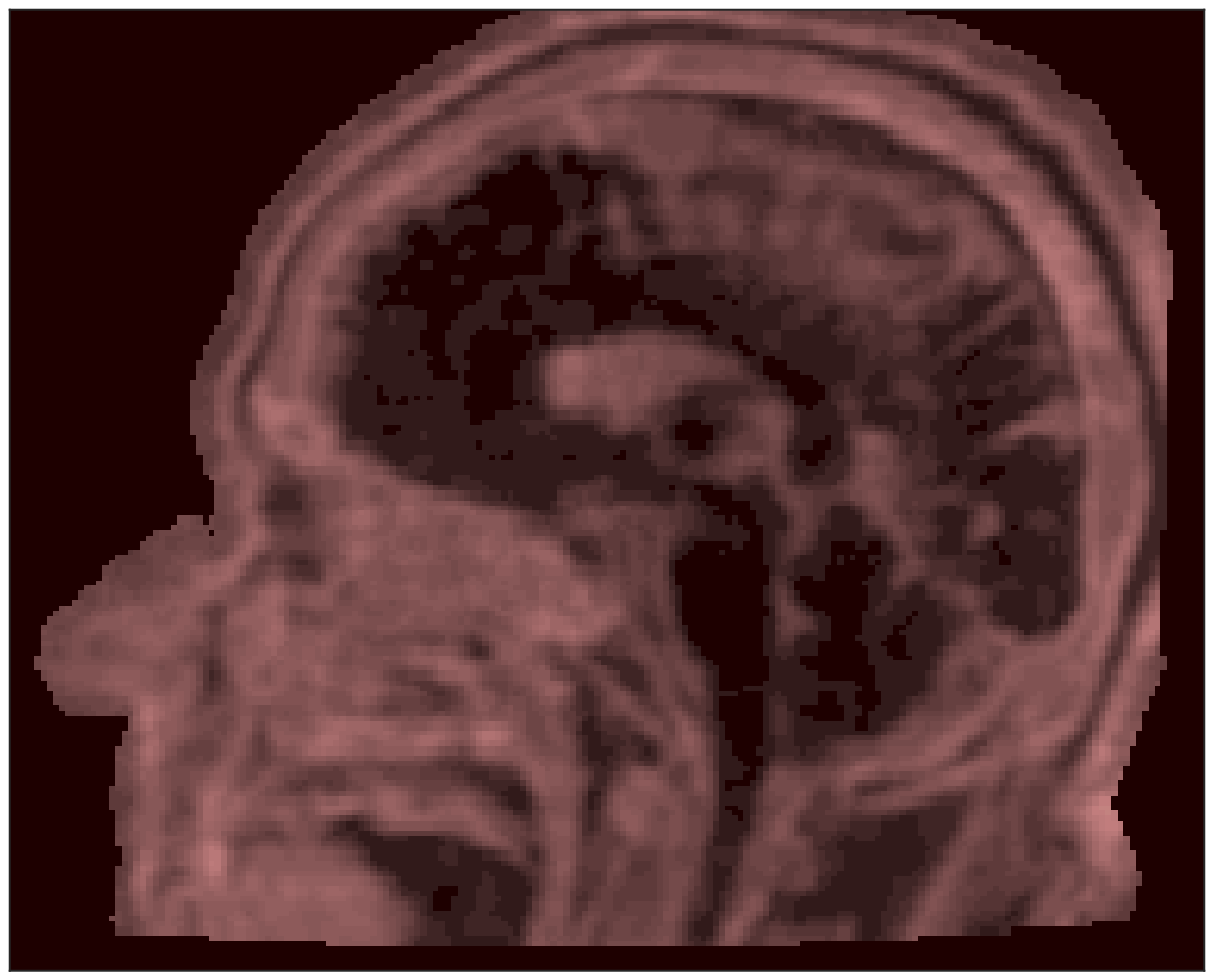}
\end{subfigure}
\\

\vspace*{0.2cm}
NIGS & 
\begin{subfigure}{0.16\linewidth}
\centering
\includegraphics[width = 0.99\textwidth, keepaspectratio]{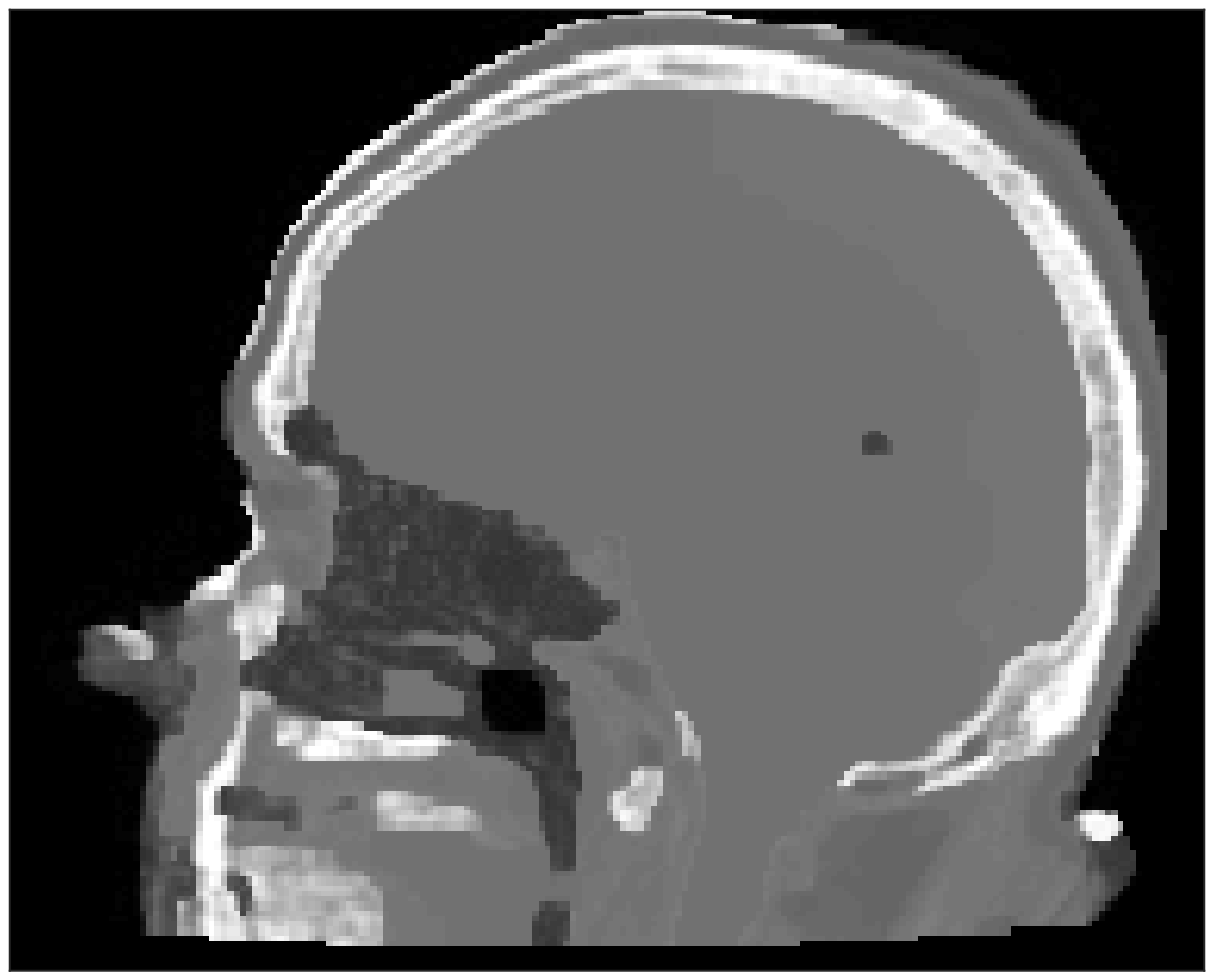}
\end{subfigure} & &
\begin{subfigure}{0.16\linewidth}
\centering
\includegraphics[width = 0.99\textwidth, keepaspectratio]{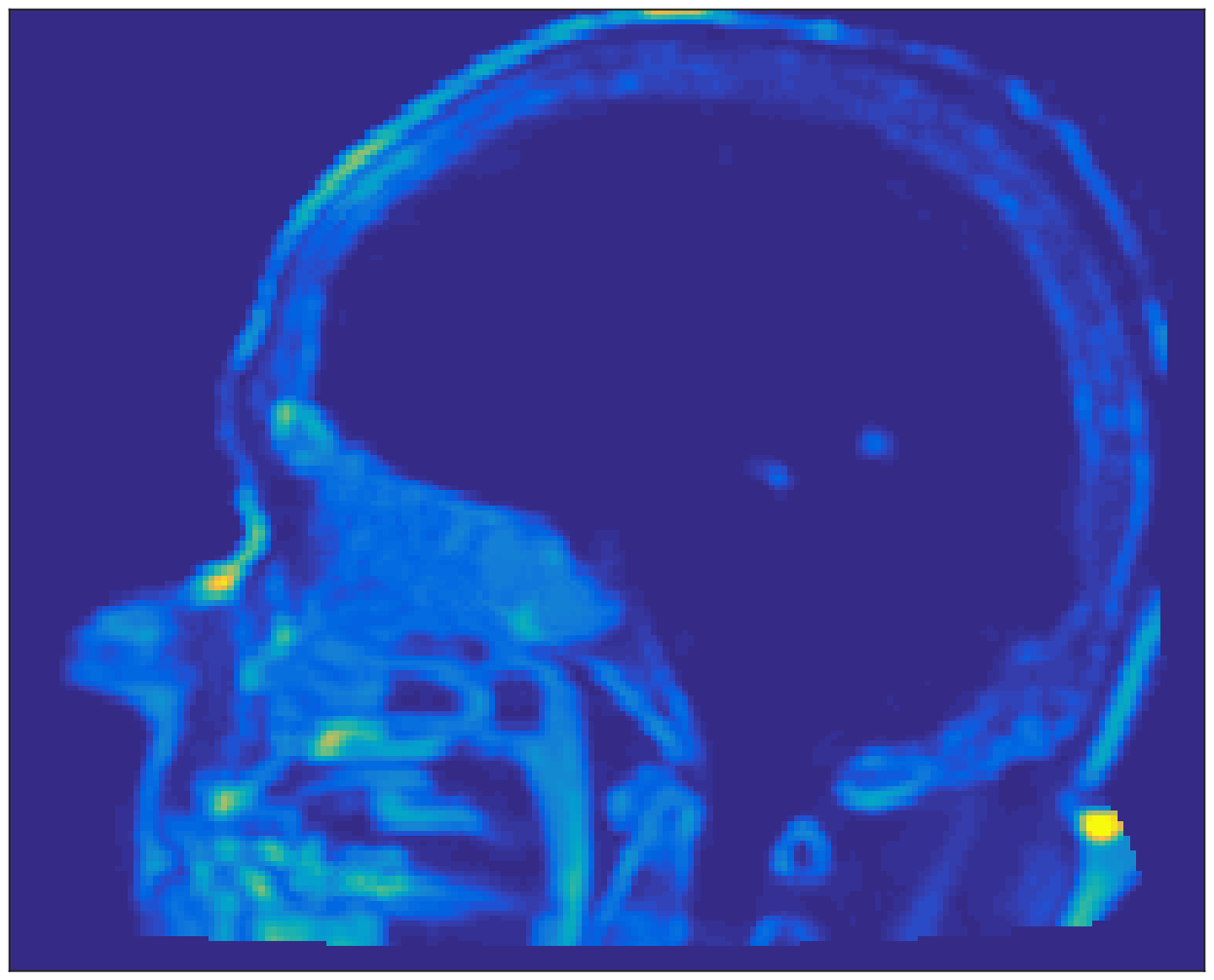}
\end{subfigure} &&
\begin{subfigure}{0.16\linewidth}
\centering
\includegraphics[width = 0.99\textwidth, keepaspectratio]{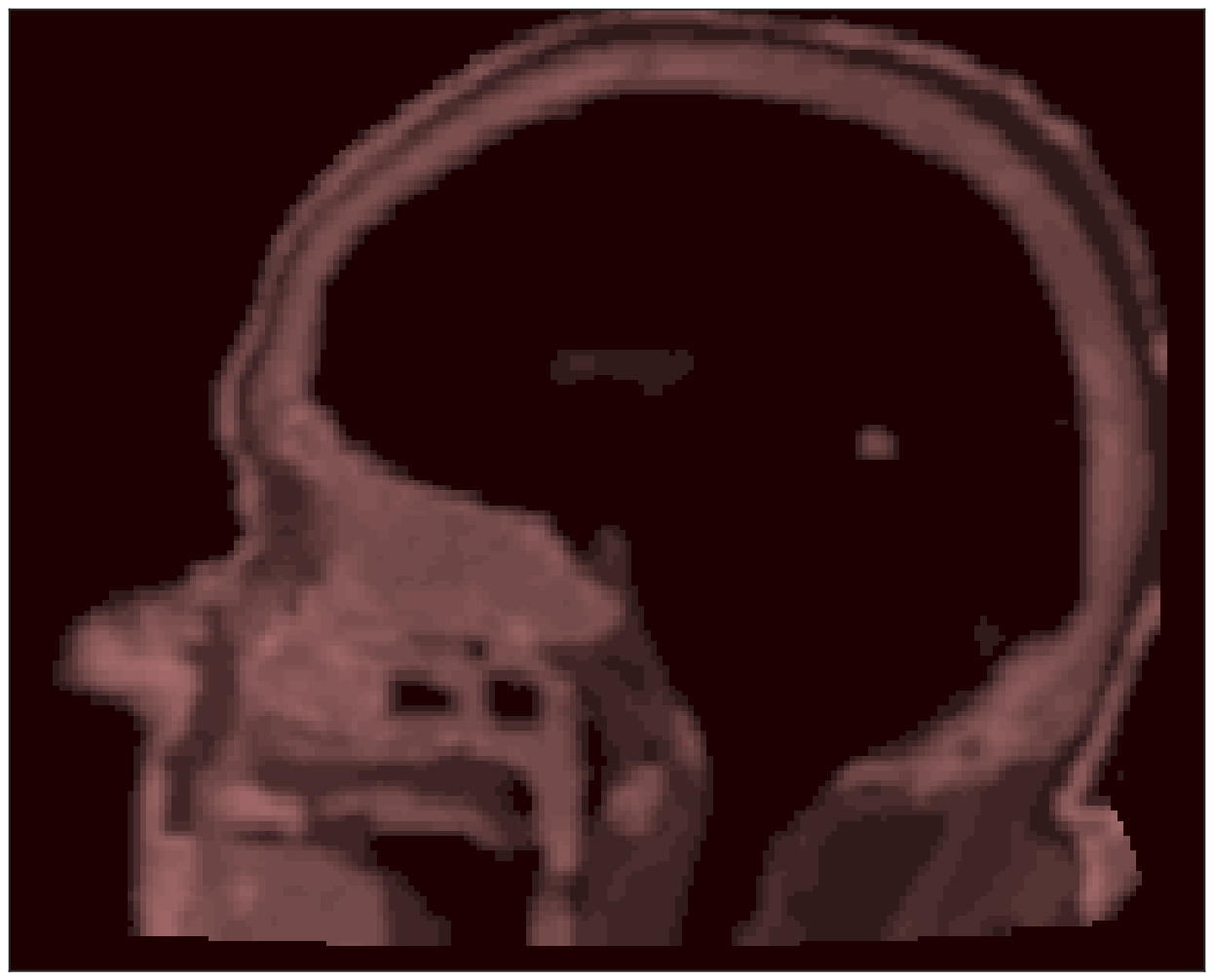}
\end{subfigure}

\end{tabular}

\caption{Generated s-CT images for one of the patients in the study. The first row shows the true CT images. The other four rows show the generated s-CT images using the four different models. The different columns show the s-CT images (left), the voxelwise absolute errors between the corresponding s-CT and true CT image (mid), and the voxelwise conditional standard deviations (right). } 
\label{fig:exampleResults}
\end{figure}

Figure \ref{fig:exampleResults} shows the true CT image together with the corresponding s-CT images for a selected patient. Since the data is three-dimensional we present it as an intersection viewed in profile of the head. The absolute errors and conditional standard deviation are also shown for comparison.
Note that the non-spatial models give more noisy predictions compared to the spatial ones. All models seem to have the most trouble predicting values in the regions where air, soft tissue and bone interacts such as in the nasal and throat cavities. These regions have short T2* values even without presence of low T2 values due to susceptibility effect \citep{lit:adamCT}, and apparently the T1 information acquired by the two flip angles is not enough to classify these regions correctly. For the spatial model these prediction problems are amplified by the fact that there are many small regions of soft tissue, air and bone close to each other. Since the spatial model allow for classes to cluster by spatial attraction, this will diminish the probability of classifying small regions close to each other to different classes as would be needed to make proper predictions in these regions. 

Apparently this clustering effect of the spatial models are advantageous overall, since otherwise the estimated models would have had $\spp$ values close to zero.

\subsection{Median filtering}
\label{sec:medianresults}
The purpose of the spatially dependent model is to increase the probability of a certain class for a voxel when this class is predominately common in the neighborhood of said voxel. This in turn will make regions of s-CT images more homogeneous while still allowing for sharp edges in between two tissue types. 

Applying a two-dimensional median filter (letting each voxel assume the median value in a defined spatial neighborhood to it) \citep{lit:Sonka} in a post-processing step to a non-spatial model would be one simple but less controlled way of giving the images these properties. A natural question is if our non-spatial model with such a median filter is comparable in prediction performance to the more complicated spatial models. In order to assess this we applied a median filter with a kernel of radius 2 (each voxel is the median of the voxel values among the nearest neighbors including the voxel itself) to the predicted images before calculating the prediction errors. This post-processing step was performed for all of the four models and Table \ref{table:optimMAEMedian} show the corresponding prediction errors. 

One can conclude that median filtering improves the s-CT images both in MAE and RMSE sense. For the non-spatial models the errors are reduced with about $5\%$ and for the spatial models with about $2\%$ in MAE (similar improvements for RMSE). Even after the median filtering step the spatial models show a considerable advantage over the non-spatial ones.

\begin{table}[t]
\centering
\caption{Prediction errors of the models at the number of classes were MAE reached its minimum for each model when using a spatial median filter on the predicted CT values in a post-processing step. The models are compared both using mean and median as predictor. "Ratios" compare the corresponding error with the reference model (GMM with mean predictor).}
\begin{tabular}{l | c | c c | c c }
Model & Classes & MAE Median & MAE Ratio & RMSE Median & RMSE Ratio \\
\hline
GMM & 9 & 168.2 & $94.3\%$ & 328.7 & $93.0 \%$ \\
GMM Median& 9 & 160.2 & $89.9 \%$ & 342.0 & $96.8 \%$\\
GMMS & 10 & 151.6 & $85.0\%$ & 316.8 & $89.6 \%$ \\
GMMS Median& 10 & 152.4 & $ 85.5 \%$& 321.2 & $90.9 \%$\\
NIG & 7 & 170.4  & $95.6 \%$ & 331.1 & $93.7 \%$ \\
NIG Median& 7 & 157.0 & $ 88.1 \%$& 339.5 & $96.1 \%$\\
NIGS & 7 & 143.8  & $80.7 \%$ & 302.4 & $85.6 \%$ \\
NIGS Median & 7 & 144.5 & $81.0 \%$& 309.9 & $87.7 \%$ \\
\hline
\end{tabular}
\label{table:optimMAEMedian}
\end{table}

\section{Discussion}\label{sec:discussion}
We have presented a class of spatially dependent mixture models that can be used to generate a three dimensional substitute CT image using information from MR images. We also introduced a computationally efficient algorithm that can perform maximum likelihood parameter estimation of the model without the need for ever evaluating the actual likelihood. This estimation method is applicable to a much larger class of problems where the likelihood is intractable to compute but where the gradient of it can be approximated.

The proposed model (NIGS) and variations of it were compared with a reference model (GMM) that has already shown promising research results. The NIGS model, which uses NIG mixture distributions and spatially dependent prior probabilities of the latent class memberships, attained the smallest prediction error measured both in MAE and RMSE. Compared to using a non-spatial GMM the prediction error decreased with $17.9\%$ in MAE and $12.7\%$ in RMSE. Using the spatially dependent model with Gaussian mixture distributions (GMMS) showed advantages to the non-spatial models but had larger prediction errors in both MAE and RMSE compared to NIGS. The spatially independent NIG mixture model (NIG) showed similar prediction performance to the original GMM model.

Compared to the work of \citet{lit:adamCT} we also evaluated the conditional median as a predictor for s-CT generation. If one is mainly interested in minimizing the MAE this is theoretically a better predictor and for a small number of mixture classes, conditional median yielded a smaller MAE than conditional mean for all models. 
For the NIGS model, the gain of decreasing the MAE by using the conditional median is declining with the number of classes and after five classes there is no apparent difference whichever predictor is used. For the GMMS, the MAE was comparable in all cases. 
The non spatial models showed a consistent advantage of using the conditional median for minimizing MAE. 
The RMSE was larger for all models when using the conditional median. Therefore we recommend using the conditional mean for predicting CT values when working with the spatial models.

The mean CRPS* values indicate that not only the point estimates but also the conditional distributions in general are more accurate for the NIGS and GMMS models. The NIG model is however consistently worse of than GMM in terms of CRPS*. This is surprising since the GMM is a limiting case of NIG, especially since it in conjunction with the spatial model clearly outperforms GMMS. 

A median filtering post processing step showed a slight advantage for all models. Both MAE and RMSE errors decreased and at most it yielded a $5.7\%$ improvement in MAE. The comparatively small gain shows that the predictive performance of the spatial model can not simply be synthesized by a median filtered non-spatial prediction. At the same time we would recommend using such a post-processing step since there are some improvements, especially to the non-spatial models. Using the NIGS model with the median filter give a $19.3\%$ improvement compared to the reference model.

All four models exhibited a decreasing marginal gain of adding further classes and using more than seven does not affect the predictive performance significantly. Since, the computational cost increases linearly with the number of classes, the parameter estimation converges slower, and there will at some point be a risk of overfitting, we suggest choosing $K = 7$ for all of the models.

The regions with the worst predictive performance seem to be the nasal/throat cavities. This is in line with prior work \citep{lit:adamCT} and is mainly due to lack of information in the MR images for the regions. The spatial model did not show any advantages in these regions and one can even argue that the model inherently counteracts enhanced prediction due to the spatially quick alternations of tissue types in these regions. If these regions are of particular importance in an application the spatial model could be enhanced by allowing for spatially varying $\beta$ parameters. One could for instance let the $\beta$ parameter be a low order polynomial of the spatial coordinates in some reference system, for instance the reference system proposed in \citet{lit:adamSpat}. Estimating $\beta$ would then correspond to estimating the coefficients of the polynomial. This can be done using the proposed gradient method and is something to include in future work.

\section*{Acknowledgements}
This work has been supported by grants from the Knut and Alice Wallenberg foundation and the Swedish Research Council Grants 2008-5382 and 340-2013-5342.

\bibliography{shell}

\appendix

\section{MCMC sampling from the latent field}\label{app:MCMC}
In Section \ref{sec:EMGradient} the posterior probability of voxel $i$ to be a member of class $k$ was needed to perform an E-step. These probabilities are equivalent to taking expectations of indicator functions over the latent field $\bs{Z}_{-i}$ conditioned on the observed data. As was stated, we can not calculate these expectations explicitly so instead we estimate them by Monte Carlo integration. In the more general case we have an expression on the form $\mathbb{E}_{\bs{Z}}[ g(Z_i, x_i , \bs{Z}_{-i} )| \bs{x}]$ for some function $g(Z_i, x_i , \bs{Z}_{-i} )$ that is explicitly available. A Monte Carlo approximation for such an expression yields
\begin{align}
\mathbb{E}_{\bs{Z}}[ g(Z_i, \bs{x}_i , \bs{Z}_{-i} ) | \bs{X} = \bs{x}] &= \sum_{\bs{z}} g(z_i,\bs{x}_i , \bs{z}_{-i}) \mathbb{P}(\bs{Z} = \bs{z}| \bs{X} = \bs{x}) \\
&\approx  \frac{1}{J} \sum_{j=1}^J g \left(\hat{z}_i^{(j)},\bs{x}_i , \hat{\bs{z}}_{-i}^{(j)} \right),
\end{align}
where $J$ is the number of realizations in the MC simulation and $\hat{\bs{z}}^{(j)}$ is the $j$:th realization of $\bs{Z} | \bs{X}$.
Using Rao-Blackwellization \citep{lit:RobertCasella} a more efficient estimation of the conditional expectation can be computed as
\begin{align}
\mathbb{E}_{\bs{Z}} &\left[ g(Z_i, \bs{x}_i , \bs{Z}_{-i} )| \bs{X} = \bs{x} \right] \approx  \frac{1}{J} \sum_{j=1}^J \sum_{k=1}^K g \left(k,\bs{x}_i , \hat{\bs{z}}_{-i}^{(j)} \right)  \mathbb{P} \left( Z_i = k \middle| \bs{X} = \bs{x}, \bs{Z}_{-i} = \hat{\bs{z}}_{-i}^{(j)} \right)
\label{eq:simExp}.
\end{align}

To be able to sample realizations of $\bs{Z} | \bs{X}$ one can note that by Bayes theorem  $\mathbb{P}(Z_i | \bs{Z}_{\bs{-i}}, \bs{X}) \propto f( \bs{X}_i | Z_i) \mathbb{P}( Z_i | \bs{Z}_{\bs{-i}} )$. Further, it is possible to sample from the full conditionals of $\bs{Z} | \bs{X}$ and Gibbs sampling \cite[Chapter 5]{lit:Winkler} can therefore be used. A blocking algorithm can be utilized to enhance the performance of the Gibbs sampling by partitioning the voxels in two mutually exclusive sets $\bs{Z} = [\bs{Z}_b, \bs{Z}_w]$ as in Figure \ref{subfig:checkers} (black and white balls). Due to the first-order neighborhood structure of the MRF model, the full conditional probabilities are only dependent on the nearest neighbors, i.e. the voxels of $\bs{Z}_b $ are independent of each other conditioned on $\bs{Z}_w$ and vice verse. A blocked Gibbs sampler can therefore be used, by iteratively sample from $\bs{Z}_b | \bs{Z}_w, \bs{x}$ and then $\bs{Z}_w | \bs{Z}_b, \bs{x}$.

The pseudolikelihood was derived by making an approximation on the joint distribution of $\bs{Z}$, see Equation \eqref{eq:pseudoLikelihood}. This approximation induces a posterior distribution $\tilde{\mathbb{P} }(\bs{Z} | \bs{X})$. 
To maximize the pesudolikelihood, the E-step in both the EM- and EM gradient-algorithm should be performed with regards to this induced probability distribution. The above sampling scheme samples from the true $\bs{Z | \bs{X}}$. In order to sample from the induced distribution one can construct importance weights from the following relation:
\begin{align}
&\tilde{\mathbb{P}} \left( \bs{Z} | \bs{X} \right) = C(\bs{\prp}, \spp) \frac{\mathbb{P}(\bs{X} | \bs{Z})}{\mathbb{P}(\bs{X})} \mathbb{P} \left( \bs{Z}_b | \bs{Z}_w \right) \mathbb{P} \left( \bs{Z}_w | \bs{Z}_b \right) = C(\bs{\prp}, \spp) \frac{ \mathbb{P} \left( \bs{Z}_w | \bs{Z}_b \right) }{\mathbb{P} \left( \bs{Z}_w  \right)} \mathbb{P} \left( \bs{Z} | \bs{X} \right) ,
\end{align}
where $C(\bs{\prp}, \spp)$ is a normalizing constant for the induced probability mass function. This gives that
\begin{align}
&\tilde{\mathbb{E}}_{\bs{Z}} \left[ g(Z_i, \bs{x_i}, \bs{Z}_{-i}) \middle| \bs{X} = \bs{x} \right] 
= \mathbb{E}_{\bs{Z}} \left[ g(Z_i, \bs{x_i}, \bs{Z}_{-i}) C(\bs{\prp}, \spp) \frac{ \mathbb{P} \left( \bs{Z}_w | \bs{Z}_b \right) }{\mathbb{P} \left( \bs{Z}_w  \right)} \middle| \bs{X} = \bs{x} \right].
\end{align}

By the use of the Hammersley-Clifford theorem \citep[Chapter 4]{lit:Winkler} the denominator of the correction factor $c(\bs{Z}) = C(\bs{\prp}, \spp) \frac{ \mathbb{P} \left( \bs{Z}_w | \bs{Z}_b \right) }{\mathbb{P} \left( \bs{Z}_w  \right)}$ is known up to a normalizing constant, see equation \eqref{eq:Gibbs}. $c(\bs{Z})$ can therefore be expressed as
\begin{align}
c(\bs{Z}) = C(\bs{\prp}, \spp) W(\bs{\prp}, \spp)  \frac{ \prod_{l \in w} \exp\left( - \sum_{m \in \mathcal{N}_l} \mathbb{I}_{z_l = z_m} \beta \right)}{  \prod_{l \in b} \left( \sum_{k=1}^K \exp\left( - \prp_{k} - \sum_{m \in \mathcal{N}_l} \mathbb{I}_{z_m = k}\beta  \right) \right) } .
\end{align}
Where $W(\bs{\prp}, \spp)$ is the unknown partition function of $\mathbb{P}(\bs{Z})$, $w$ is the set of all voxels marked as "white balls" and $b$ is the set of all voxels marked as "black balls" from the blocking scheme, see Figure \ref{subfig:checkers}.

By utilizing self-normalizing importance sampling \citep[Chapter 3]{lit:Robert} it is possible to approximate the expectation by
\begin{align}
\tilde{\mathbb{E}}_{\bs{Z}} \left[ g(Z_i, \bs{x_i}, \bs{Z}_{-i}) \middle| \bs{X} = \bs{x} \right] 
\approx \sum_{k}^K \frac{ \sum_{j=1}^J  g(k, \bs{x_i}, \hat{\bs{z}}^{(j)}_{-i}) c\left(k, \hat{\bs{z}}_{-i}^{(j)}\right)  \mathbb{P}\left( Z_i = k | \bs{X} = \bs{x}, \bs{Z}_{-i} = \hat{\bs{z}}_{-i}^{(j)} \right) }{ \sum_{j = 1}^J  c\left(k, \hat{\bs{z}}_{-i}^{(j)}\right) }
\end{align}

However, approximating $\tilde{\mathbb{E}}[...]$ satisfactory using self normalizing importance sampling is more computationally demanding and the resulting parameter estimates are in practice very close to the ones obtained by just approximating $\tilde{\mathbb{E}}_{\bs{Z}}[...] \approx \mathbb{E}_{\bs{Z}}[...]$. Therefore, we use this approximation in favor of the importance sampling described above, as has been done before in similar problems \citep{lit:Celeux, lit:Rongjing}.

The computational complexity of approximating $\mathbb{E}_{\bs{Z}}[ g(Z_i, x_i , \bs{Z}_{-i} )| \bs{x}]$ for a general function $g$ is hence of order $\mathcal{O}(J N K^2 )$, where $J$ is the number of Monte Carlo iterations, $K$ is the number of classes and $N$ is the number of voxels in the image. The $N$ and one of the $K$ factors are attributed to a Gibbs sampling stage. The $J$ and the other $K$ factor are attributed to the sum in the Rao-Blackwellisation of the Monte Carlo integration. Note that approximating the posterior probabilities of the latent field only has a complexity of $\mathcal{O}(J N K )$ since the sum over the classes in \eqref{eq:simExp} reduces to the term involving $k = z_i$ due to the indicator function.

For a valid MCMC simulation $J$ needs to be large enough so that the burn in phase of the Gibbs simulation can be omitted. However, the EM gradient algorithm is an iterative procedure and in each iteration one needs new evaluations of $\pi(z_i | \bs{x})$. In order to reduce the computations one can make use of the iterative scheme of the EM gradient algorithm. 
$J$ can still be chosen small since each gradient iteration feed off the MCMC sampling of the former one. The idea behind this is that the spatial field characterized by the spatial parameters ($\prp_k, \spp$) will be reasonably similar between two consecutive gradient iterations. One can then use the field generated in the former gradient iteration as an initial value for the new MCMC simulation. Through this trick the need for a burn in period is basically eliminated.
In our implementation we used $J = 10$ for the parameter estimation phase.

\section{EM Gradient conditional line search}\label{app:condLineSearch}
Performing a conditional line search in the EM gradient algorithm corresponds to numerically finding a value of $\delta$ that maximizes 
\begin{align}
Q \left(  \Theta^{(j)} + \delta H^{-1}\left(\log \tilde{L}\left(\Theta^{(j)} ; \bs{x} \right) \right) \nabla \left(\log \tilde{L}\left(\Theta^{(j)} ; \bs{x} \right) \right)   \middle| \bs{X} = \bs{x}; \Theta^{(j)} \right) ,
\end{align}
where 
\begin{align}
Q &\left(\Theta | \Theta^{(j)} \right) = \sum_{i = 1}^N \sum_{k = 1}^K  \log f(x_i| Z_i = k; \Theta) \mathbb{P}\left( Z_i = k | \bs{X} = \bs{x} ; \Theta^{(j)} \right) \\
&+ \sum_{i = 1}^N \sum_{k = 1}^K  \mathbb{E}_{\bs{Z}_{-i}} \left[ \log \pi( Z_i = k | \mathbf{Z}_{-i} ; \Theta) \middle| \bs{X} = \bs{x}, \Theta^{(j)} \right]  \mathbb{P}\left( Z_i = k |\bs{X} = \bs{x} ; \Theta^{(j)} \right) .
\label{eq:condloglik}
\end{align}
A line search requires evaluations of the function $Q$ for several values of $\delta$. Fortunately, both $\log f(x_i| Z_i = k; \Theta)$ and $\mathbb{E}_{\bs{Z}_{-i}} \left[ \log \mathbb{P}( Z_i = k | \mathbf{Z}_{-i} ; \Theta) \middle| \bs{X}=\bs{x}, \Theta^{(j)} \right]$ can be calculated explicitly within a feasible computational cost. This since the first term does only depends on the current voxel $i$ and the second term does only need to be summed over all possible states of $Z_j$ for the nearest neighbors to voxel $i$ since the $Z$-field is Markov. This is feasible since both the number of classes and the number of neighbors are typically small.  Also, the probabilities $\mathbb{P}( Z_i = k | \bs{X}=\bs{x} ; \Theta^{(j)})$, are already approximated with the Monte Carlo simulation in the gradient step and can be reused without further computations, hence the line search is a tractable method to ensure convergence. 

Remember the equation
\begin{equation}
\begin{split}
\nabla \log &\tilde{L}( \Theta ; \bs{x}) =  \sum_{i=1}^N\mathbb{E}_{\mathbf{Z}} \left[   \nabla \log f( \bs{x}_i | Z_i ; \Theta ) \middle| \bs{X} = \bs{x} ; \Theta \right]  \\
&+  \sum_{i=1}^N \mathbb{E}_{\mathbf{Z}} \left[\nabla \log \mathbb{P}( Z_i = z_i | \bs{Z}_ {-i} = \mathbf{z}_{-i} ; \Theta )  \middle| \bs{X} = \bs{x} ; \Theta \right] 
\end{split}
\end{equation}
from Section \ref{sec:EMGradient}.
The two terms inside the expectation depends on two mutually exclusive sets of parameters. The first term is associated with the parameters of the mixture distributions ($\lkp_k$, $\skp_k$, $\kp_k$, $Q_k$), and the second term is associated with the parameters of the MRF ($\bs\spp$ and $\bs\prp$). Conditioned on $\bs{z}$, the gradient with regard to the MRF parameters is not dependent on the mixture distributions parameters and vice versa. Hence, the approximate M-step of the EM gradient method can be separated into two separate Newton steps, one for the MRF parameters and one for the mixture parameters. It is therefore reasonable to choose the step lengths separately for the two steps. This can be beneficial since one set of parameters might need a smaller step size in some regions of the parameter space while constraining the other parameter set to the same small step size might inhibit convergence speed.   

We take advantage of this separation by performing a line search for the parameters of the mixture distributions, which often has shown to need a smaller step size than the one proposed by the approximate Newtons method. However, we simply use a fixed step length for the MRF parameters since the proposed step lengths of the approximate Newtons method seem to be satisfactory. Also the amount of indexing needed to implement the line search in this case significantly increases the execution time of the parameter estimation.

\section{Derivations of properties of the NIG distribution}
\label{app:propNIG}

In order to prove Propositions \ref{prop:marginalNIG} and \ref{prop:conditionalNIG}, let us recall some properties of the NIG distribution. If $\mathbf{X}$ is NIG distributed, we have that $\mathbf{X}|V \sim \pN(\lkp - \bs{\skp} v,vQ^{-1})$, where $V\sim\pIG(\tau,\sqrt{\tau/2})$. The density of $\mathbf{X}$ can thus be derived by computing 
\begin{align}\label{eq:NIGintegral}
f(\bs{x}) = \int f(\bs{x},v) dv = \int f(\bs{x}|v)f(v) dv
\end{align}
where $f(\bs{x}|v)$ is the density of $\mathbf{X}|V$ and $f(v)$ is the density of the inverse Gaussian distribution, which is shown in Appendix \ref{app:GIG}. It is easy to evaluate the integral if one identify the factors including $v$ as a probability density function of a GIG distribution with parameters $\nu = -\frac{d+1}{2}$, $a = \bs{\skp}^T Q \bs{\skp} + \frac{\kp}{\delta^2}$, $b = (\bs{x}-\bs{\lkp})^TQ(\bs{x}-\bs{\lkp}) + \kp $ and without correct normalizing constant.

\subsection{Proof of Proposition \ref{prop:marginalNIG}}
\label{sec:appMarginal}
We have that
\begin{align}
f \left( \bs{x}^B \right) &= \int f \left( \bs{x}^B | v \right) f(v) dv = \int \left( \int f(\bs{x} | v)d\bs{x}^A \right) f(v) dv  \\
f \left( \bs{x}^B | v \right) &=  \frac{\sqrt{|\hat{Q}|}}{(2\pi v)^{(d-d_m)/2} } e^{  -\frac{1}{2v} \left( \bs{x}^B - \bs{\lkp}^B - \bs{\skp}^B v \right)^T \hat{Q} \left( \bs{x}^B - \bs{\lkp}^B - \bs{\skp}^B v \right)  } .  
\end{align}
The resulting density is now acquired by recognizing the integral as the integral over a GIG distribution without a proper normalization analogous to how the density for the joint distribution was acquired by the integral \eqref{eq:NIGintegral}. This gives

\begin{align}
f(\bs{x}^B) &= \frac{ \sqrt{ \kp |\hat{Q}|} }{ (2\pi)^{(d-d_m+1)/2}  } e^{\left(  \left( \bs{x}^B - \bs{\lkp}^B \right)^T \hat{Q}  \bs{\skp}^B + \frac{\kp}{\delta}  \right)} 2 K_{\hat{\nu}}(\sqrt{\hat{a}\hat{b}}) \left( \frac{\hat{b}}{\hat{a}} \right)^{\frac{\hat{\nu}}{2}},
\end{align}
where the hat denotes parameters of the marginal distribution and relates to the original parameters as follows: $\hat{a} = \bs{\skp}^{B} \hat{Q} \bs{\skp}^{B} + \frac{\kp}{\delta^2},
\hat{b} = \left( \bs{x}^{B} - \bs{\lkp}^{B} \right)^T \hat{Q} \left( \bs{x}^{B} - \bs{\lkp}^{B} \right) + \kp, \hat{\nu} = -\frac{d-|A|+1}{2},$ and $\hat{Q} = \left( \Sigma^{BB} \right)^{-1}$. We can identify this as the NIG distribution given in the proposition. \qed

\subsection{Proof of proposition \ref{prop:conditionalNIG}}
We have that
\begin{align}
f(\bs{x}^A | \bs{x}^{B}) &= \int f(\bs{x}^A | \bs{x}^{B}, v)f(v | \bs{x}^{B}) dv
\end{align}
where $f( v| \bs{x}^{B} ) \propto f( \bs{x}^{B} | v )f(v)$ and 
$$
\bs{x}^A | \bs{x}^{B}, v \sim \pN \left( \bs{\lkp}^A + \bs{\skp}^A v - \left(Q^{AA}\right)^{-1}Q^{AB}(\bs{x}^{B} - \bs{\lkp}^{B} - \bs{\skp}^{B}v) , v\left(Q^{BB}\right)^{-1}   \right) 
$$
From Section \ref{sec:appMarginal} we know that $f( \bs{x}^{B} | v )f(v)$ corresponds to the density function of a GIG distribution without normalization with parameters ($\hat{\nu}$, $\hat{a}$, $\hat{b}$). Hence $v | \bs{x}^{B} \sim \text{GIG}(\hat{\nu}$, $\hat{a}$, $\hat{b})$ and since $\bs{X^A}| \bs{X^B}, v $ is Gaussian distributed with mean $ \bs{\lkp}^A + \bs{\skp}^A v - \left(Q^{AA}\right)^{-1}Q^{AB} ( \bs{x}^{B} - \bs{\lkp}^{B} - \bs{\skp}^{B} v )  := \bs{X}^A - \bs{L} - \bs{O}v$ and precision matrix $\frac{1}{v}Q^{AA}$ we get:
\begin{align}
f &\left(\bs{x}^A | \bs{x}^B \right) = \int \frac{\sqrt{|Q^{AA}|}}{(2\pi v)^{|A|/2}} \left( \frac{\hat{a}}{\hat{b}} \right)^{\hat{\nu}/2} \frac{ \exp\left(-\frac{1}{2v} \left( \bs{L} + \bs{O}v \right)^T Q_{AA} \left( \bs{L} + \bs{O}v \right) -\frac{\hat{a}v}{2} - \frac{\hat{b}}{2v} \right)   }{2K_{\hat{\nu}} \left( \sqrt{\hat{a}\hat{b}} \right) }   v^{\hat{\nu}-1}  dv \\
&\propto  \int v^{\hat{\nu}-1 - |A|/2} \exp\left(-\frac{1}{2v} \left(\bs{L} + \bs{O}v \right)^T Q^{AA} \left( \bs{L} + \bs{O}v \right) -\frac{\hat{a}v}{2} - \frac{\hat{b}}{2v} \right)    dv \\
&\propto e^{-\bs{L}^T Q^{AA} \bs{O}}  \int v^{\hat{\nu}-1 - |A|/2} \exp\left( - \frac{v}{2} \left( \bs{O}^T Q^{AA} \bs{O} + \hat{a}\right) -\frac{1}{2v}\left( \bs{L}^T Q^{AA}\bs{L} + \hat{b} \right)  \right)     dv .
\end{align}
The integral can be identified as a integral over a GIG distribution without proper normalization. This give us the stated conditional density function. Moreover, 
\begin{align}
&\bs{X}^A | \bs{X}^B \sim \pGH(\tilde{\bs{\lkp}}, Q^{AA}, \tilde{\bs{\skp}}, \hat{\nu}, \hat{a}, \hat{b}),
\end{align}
with the definition of a generalized hyperbolic distribution as in \ref{app:GH}.
Here, 
$\tilde{\bs{\lkp}} =  \bs{\lkp}^A - \left(Q^{AA}\right)^{-1}Q^{AB} \left(\bs{x}^B - \bs{\lkp}^B \right)$ and $\tilde{\bs{\skp}} = \bs{\skp}^A + \left(Q^{AA}\right)^{-1}Q^{AB}\bs{\skp}^B$.

From this the conditional expectation can be derived as
\begin{align}
\mathbb{E} &[\bs{X}^A | \bs{x}^B] = \int \bs{x}^A f(\bs{x}^A | \bs{x}^B) d\bs{x}^A = \int \bs{x}^A \int f(\bs{x}^A| \bs{x}^B, v)f(v | \bs{x}^B) dv d\bs{x}^A  \\
&= \int \mathbb{E}[\bs{X}^A | \bs{x}^B, v] f(v| \bs{x}^B) dv  \\
&= \int \left[ \bs{\lkp}^A + \bs{\skp}^A v - \left(Q^{AA}\right)^{-1}Q^{AB} \left(( \bs{x}^B - \bs{\lkp}^B - \bs{\skp}^B v \right)  \right] f(v| \bs{x}^B) dv  \\
&= \left[ \bs{\lkp}^A - \left(Q^{AA}\right)^{-1} Q^{AB} \left( \bs{x}^B - \bs{\lkp}^B \right) \right] + \left[ \bs{\skp}^A + \left(Q^{AA}\right)^{-1}Q^{AB}\bs{\skp}^B \right] \mathbb{E}[v|\bs{x}^B ]  \\
&= \left[ \bs{\lkp}^A - \left(Q^{AA}\right)^{-1}Q^{AB} \left( \bs{x}^B - \bs{\lkp}^B \right) \right] \\
&+ \left[ \bs{\skp}^A + \left(Q^{AA}\right)^{-1}Q^{AB}\bs{\skp}^B \right] \sqrt{\frac{\hat{b}}{\hat{a}}} \frac{ K_{\hat{\nu}+1}(\sqrt{\hat{a}\hat{b}}) }{K_{\hat{\nu}}(\sqrt{\hat{a}\hat{b}})}  = \tilde{\bs{\lkp}} + \tilde{\bs{\skp}} \sqrt{\frac{\hat{b}}{\hat{a}}} \frac{ K_{\hat{\nu}+1}(\sqrt{\hat{a}\hat{b}}) }{K_{\hat{\nu}}(\sqrt{\hat{a}\hat{b}})},
\end{align}
where second to last step used the expression for the expected value of a GIG variable, given in Appendix \ref{app:GIG}. Similarly the conditional covariance can be derived by
\begin{align}
\mathbb{E} &\left[ \left(\bs{X}^A\right)^2 | \bs{x}^B \right] = \int \left[ \Cov(\bs{X}^A | \bs{x}^B, v) + \mathbb{E}[\bs{X}^A | \bs{x}^B, v] \mathbb{E}[\bs{X}^A | \bs{x}^B, v]^T \right] f(v | \bs{x}^B) dv  \\
&= \int \left[ v \left( Q^{AA} \right)^{-1} + (\tilde{\bs{\lkp}} + \tilde{\bs{\skp}}v) (\tilde{\bs{\lkp}} + \tilde{\bs{\skp}}v)^T \right] f(v | \bs{x}^B) dv  \\
&= \int \left[ \tilde{\bs{\lkp}}\tilde{\bs{\lkp}}^T + \left( \tilde{\bs{\lkp}}\tilde{\bs{\skp}}^T + \tilde{\bs{\skp}}\tilde{\bs{\lkp}}^T +  \left(Q^{AA}\right)^{-1} \right) v + \tilde{\bs{\skp}}\tilde{\bs{\skp}}^T v^2 \right] f(v | \bs{x}^B) dv  \\
&= \tilde{\bs{\lkp}}\tilde{\bs{\lkp}}^T + \left( \tilde{\bs{\lkp}}\tilde{\bs{\skp}}^T + \tilde{\bs{\skp}}\tilde{\bs{\lkp}}^T + \left(Q^{AA}\right)^{-1} \right) \mathbb{E}[V | \bs{x}^B] + \tilde{\bs{\skp}}\tilde{\bs{\skp}}^T \mathbb{E}[V^2 | \bs{x}^B]  \\
&= \tilde{\bs{\lkp}}\tilde{\bs{\lkp}}^T + \left( \tilde{\bs{\lkp}}\tilde{\bs{\skp}}^T + \tilde{\bs{\skp}}\tilde{\bs{\lkp}}^T + \left(Q^{AA}\right)^{-1} \right) \sqrt{\frac{\hat{b}}{\hat{a}}} \frac{ K_{\hat{\nu}+1}(\sqrt{\hat{a}\hat{b}}) }{K_{\hat{\nu}}(\sqrt{\hat{a}\hat{b}})}  + \tilde{\bs{\skp}}\tilde{\bs{\skp}}^T \frac{\hat{b}}{\hat{a}} \frac{K_{\hat{\nu}+2}(\sqrt{\hat{a} \hat{b}})}{K_{\hat{\nu}}(\sqrt{\hat{a} \hat{b}})}
\end{align}
where the  expression for $\mathbb{E}[V^2 | \bs{x}^B] $ is taken from Appendix \ref{app:GIG}.
\begin{align}
\mathbb{C} &( \bs{X}^A | \bs{x}^B ) = \mathbb{E} \left[ \left(\bs{X}^A\right)^2 | \bs{x}^B \right] - \mathbb{E}[\bs{X}^A | \bs{x}^B]\mathbb{E}[\bs{X}^A | \bs{x}^B]^T  \\
&= \left[ \tilde{\bs{\lkp}}\tilde{\bs{\lkp}}^T + 
\left( \tilde{\bs{\lkp}}\tilde{\bs{\skp}}^T + \tilde{\bs{\skp}}\tilde{\bs{\lkp}}^T + \left(Q^{AA}\right)^{-1} \right) \sqrt{\frac{\hat{b}}{\hat{a}}} \frac{ K_{\hat{\nu}+1}(\sqrt{\hat{a}\hat{b}}) }{K_{\hat{\nu}}(\sqrt{\hat{a}\hat{b}})}  + 
\tilde{\bs{\skp}}\tilde{\bs{\skp}}^T \frac{\hat{b}}{\hat{a}} \frac{K_{\hat{\nu}+2}(\sqrt{\hat{a} \hat{b}})}{K_{\hat{\nu}}(\sqrt{\hat{a} \hat{b}})} \right] \\
& \quad - \left[ \tilde{\bs{\lkp}}\tilde{\bs{\lkp}}^T + 
( \tilde{\bs{\skp}}\tilde{\bs{\lkp}}^T + \tilde{\bs{\lkp}}\tilde{\bs{\skp}}^T  ) \sqrt{\frac{\hat{b}}{\hat{a}}} \frac{ K_{\hat{\nu}+1}(\sqrt{\hat{a}\hat{b}}) }{K_{\hat{\nu}}(\sqrt{\hat{a}\hat{b}})} + 
\tilde{\bs{\skp}}\tilde{\bs{\skp}}^T \left( \frac{\hat{b}}{\hat{a}} \frac{ K_{\hat{\nu}+1}(\sqrt{\hat{a}\hat{b}}) }{K_{\hat{\nu}}(\sqrt{\hat{a}\hat{b}})} \right)^2 \right]  \\
&=  \left(Q^{AA}\right)^{-1} \sqrt{\frac{\hat{b}}{\hat{a}}} \frac{ K_{\hat{\nu}+1}(\sqrt{\hat{a}\hat{b}}) }{K_{\hat{\nu}}(\sqrt{\hat{a}\hat{b}})} + \tilde{\bs{\skp}}\tilde{\bs{\skp}}^T \left[ \frac{\hat{b}}{\hat{a}} \frac{K_{\hat{\nu}+2}(\sqrt{\hat{a} \hat{b}})}{K_{\hat{\nu}}(\sqrt{\hat{a} \hat{b}})} - \left( \frac{\hat{b}}{\hat{a}} \frac{ K_{\hat{\nu}+1}(\sqrt{\hat{a}\hat{b}}) }{K_{\hat{\nu}}(\sqrt{\hat{a}\hat{b}})} \right)^2 \right].
\end{align}

The desired result is now obtained by using the equality 
$$\frac{\nu}{z}K_{\nu}(z) - K_{\nu + 1}(z) = -\frac{\nu}{z}K_{\nu}(z) - K_{\nu - 1}(z)$$
for the modified Bessel function $K_{\hat{\nu}+2}(\sqrt{\hat{a} \hat{b}})$.  \qed

\section{Distributions}

\subsection{The generalized inverse Gaussian distribution}\label{app:GIG}
A random variable $X$ has a GIG distribution if it has probability density function
\begin{align}
f(x) &= \left( \frac{a}{b} \right)^{\nu/2} \frac{x^{\nu-1}}{2K_{\nu} \left( \sqrt{ab} \right) } \exp \left( -\frac{ax}{2} - \frac{b}{2x} \right) .
\end{align}
The following expectations holds true for the GIG distribution,
\begin{align}
\mathbb{E}[ X ]  &= \sqrt{\frac{b }{a}} \frac{K_{\nu+1}(\sqrt{ba})}{K_{\nu}(\sqrt{ba})}    \\
\mathbb{E} \left[ X^{2}   \right]  &=  \frac{b}{a} \frac{K_{\nu+2}(\sqrt{a b})}{K_{\nu}(\sqrt{ab})}
\end{align}

\subsection{The inverse Gaussian distribution}
The IG is a special case of a GIG distribution with parameters $\nu = -\frac{1}{2}$, $a = \frac{\kp}{\delta^2}$, $b = \kp$, and thus has density 
\begin{align}
f(x) &= \frac{\sqrt{\kp}}{\sqrt{2\pi x^3}} \exp \left( -\frac{\kp (x - \delta)^2 }{2\delta^2 x} \right).
\end{align}

\subsection{The generalized hyperbolic distribution}\label{app:GH}
The generalized hyperbolic distribution (GH) is defined similarly to the NIG distribution \citep{lit:Hammerstein}, but instead of letting the latent variance variable be IG distributed it is instead GIG distributed.
\begin{equation}
\mathbf{X} \sim  \pGH(\boldsymbol\lkp, Q, \boldsymbol\skp, \nu, a, b) \text{ if } \begin{aligned}
\begin{cases}
&\mathbf{X} = \boldsymbol\lkp + \boldsymbol\skp V +  \sqrt{V}Q^{-\frac{1}{2}} \boldsymbol Z   \\
&V \sim \pGIG ( \nu , a, b ) \\
&\mathbf{Z} \sim \pN( \boldsymbol{\lkp} = \bs{0}, \bs{I}) 
\end{cases}
\end{aligned}
\label{eq:GHDefined}
\end{equation}

\section{Continuous Ranked Probability Score}\label{app:CRPS}
The Continuous Ranked Probability Score (CRPS) is a scoring rule that assesses how well a continuous probability distribution explains observed data. It is a proper scoring rule, i.e. the expected score is largest for the distribution from which the data actually was sampled from. It is defined as
\begin{align}
CRPS(F, x) =& -\int_{\infty}^{\infty} \left( F(y) - \mathbb{I}(y \ge x) \right)^2 dy, 
\end{align}
where $F$ is the cumulative distribution function of a distribution and $x$ is some observation. Often CRPS is used negatively oriented (CRPS* = $-$CRPS) since then it is positive and a small value close to zero corresponds to a good fit.

\citet{lit:gneitingCRPS} showed that the CRPS can be expressed as 
\begin{align}
CRPS(F, x) =& \frac{1}{2} \mathbb{E} \left[ |Y - Y'| \right] - \mathbb{E} \left[ |Y - x| \right], 
\end{align}
where $Y$ and $Y'$ are i.i.d. with a distribution that corresponds to $F$.

If $F$ corresponds to a normal distribution, then both $|Y - Y'|$ and $|Y-x|$ are distributed as a folded normal distribution. If $Y \sim \mathcal{N}(\mu, \sigma^2)$, then one has
\begin{align}
\mathbb{E} [|Y|] = 2\sigma \phi \left(\frac{\mu}{\sigma}\right) + \mu \left( 2\Phi \left( \frac{\mu}{\sigma} \right)  - 1  \right).
\label{eq:foldedExpectation}
\end{align}

In this work we need to compute the CRPS for mixture models with the class membership field $Z$ and the conditional distribution $X^A_i | \bs{X^B} = \bs{x}^b$. Hence
\begin{align}
CRPS(F, x^A_i) =& \mathbb{E} \left[  \frac{1}{2} \mathbb{E} \left[ |X_i^A - X^{A'}_i| \middle| Z_i \right] - \mathbb{E} \left[ |X^A_i - x^A_i| \middle| Z_i \right] \middle| \bs{X}^{B} = \bs{x}^B \right] \\
&= \sum_{k = 1}^K \left(  \frac{1}{2} \mathbb{E} \left[ |X^A_i - X^{A'}_i| \middle| Z_i \right] - \mathbb{E} \left[ |X^A - x^A_i| \middle| Z_i \right] \right) \mathbb{P} \left(Z_i = k \middle| \bs{X}^B = \bs{x}^B \right),
\end{align}
The posterior class probabilities, $\mathbb{P} \left(Z_i = k \middle| \bs{X}^B = \bs{x}^B \right)$, are already approximated by MCMC simulation during the s-CT prediction, see Appendix \ref{app:MCMC}. For the GMM and GMMS models the CRPS can therefore be computed explicitly given those probabilities and equation \eqref{eq:foldedExpectation} since conditioned on $Z_i$, $X^A_i | \bs{X}^B$ is normally distributed.

For the NIG mixtures we have no explicit expression of the CRPS and instead we need to compute them by Monte Carlo simulations. Just as in \citet{lit:bolin} the variance of the MC simulation can be significantly decreased by realizing that conditioned on the class and the variance variable, $V_i$, the NIG distribution is also normally distributed. One therefore need to Monte Carlo simulate the variance variable, but not the entire NIG variable, since the CRPS value of a normally distributed variable could be acquired analytically. Thus,
\begin{align}
CRPS(F, x^A_i) &= \frac{1}{2n}\sum_{k=1}^K \sum_{j=1}^n \left( \mathbb{E} \left[ |X^A_i - X^{A'}_i| \middle| V_i = v_j, V'_i = v'_j, Z_i = k\right] \right. \\
& \left. - \mathbb{E} \left[ |X^A_i - x^A_i| \middle| V_i = v_j, Z_i = k \right]  \middle| Z_i=k  \right) \cdot \mathbb{P} \left(Z_i = k \middle| \bs{X}^B = \bs{x}^B \right),
\end{align}
where $v_j$ and $v'_j$ are sampled from the current $V_i | Z_i$.

\end{document}